\RequirePackage{fix-cm}
\documentclass[twocolumn,epjc3]{svjour3}
\smartqed  
\RequirePackage{graphicx}
\RequirePackage[numbers,sort&compress]{natbib}
\usepackage{latexsym,amsmath,amsfonts,amssymb,bm,empheq,nicematrix}   
\usepackage{nicefrac} 
\usepackage{multirow}
\usepackage{tabularx}   
\usepackage{multirow}
\usepackage{nicefrac}
\usepackage{tikz}   
\usepackage{tikz-feynman}
\usetikzlibrary{shapes,arrows,positioning,automata,backgrounds,calc,er,patterns}
\tikzfeynmanset{compat=1.0.0}
\usepackage{upgreek}

\usepackage{subfigure}  

\newcommand{\ee}{\mathrm{e}}

\newcommand{\ii}{\mathrm{i}}
\newcommand{\Tr}{\mathrm{Tr}}

\newcommand{\rr}{{\text r}}
\newcommand{\PT}{\mathbb{P}^{\rm{T}}}
\usepackage{hyperref}
\hypersetup{
    colorlinks=true,
}
\begin{document}

\title{Dense polar active fluids in a disordered environment}

\author{Riccardo~Ben~Al\`i~Zinati\thanksref{e1,addr1}\,,~Marc~Besse\thanksref{e2,addr1}\,,~Gilles~Tarjus\thanksref{e3,addr1}\,,~Matthieu~Tissier\thanksref{e4,addr1}
}

\thankstext{e1}{riccardo.baz@pm.me}
\thankstext{e2}{besse@lptmc.jussieu.fr}
\thankstext{e3}{tarjus@lptmc.jussieu.fr}
\thankstext{e4}{tissier@lptmc.jussieu.fr}

\institute{Sorbonne University, CNRS-UMR7600, Laboratoire de Physique Th\'eorique de la Mati\`ere Condens\'ee,  F-75005,  Paris, France\label{addr1}
}

\maketitle

\begin{abstract}
We examine the influence of quenched disorder on the flocking transition of dense polar active matter. We consider incompressible 
systems of active particles with aligning interactions under the effect of either quenched random forces or random dilution. The system displays a continuous 
disorder-order (flocking) transition, and the associated scaling behavior is described by a new universality class which is controlled by a quenched Navier-Stokes fixed 
point. We determine the critical exponents through a perturbative renormalization group analysis. We show that the 
two forms of quenched disorder, random force and random mass (dilution), belong to the same universality class, in contrast with the situation at equilibrium.
\end{abstract}

\section{Introduction}
At almost every scale in Nature, the collective motion of large groups of individuals takes place in heterogeneous environments.
Self-propelled particles, such as bacteria, can experience the roughness and irregularity of the substrate they are moving on, as much as the motion of birds and fish can be 
influenced by the presence of topographical features such as hurdles and barriers. Collective motion of active matter results from the competition between the tendency of individuals to follow the trajectory of their neighbors, {\it i.e.}, to align, and some form of misalignment induced by noise \cite{Vicsek2012review}.
The latter can be due to time-dependent (\emph{thermal}) disturbances or to time-independent (\emph{quenched}) ones that may represent the presence of impurities or 
obstacles in the background landscape.

Numerous experimental and theoretical studies have provided a picture of the collective motion in the presence of dirty landscapes \cite{Chepizhko2013,Peruani2018,Das2018,Sandor2017}.
We focus here on dry polar active matter.
In the absence of quenched disorder, polar active systems may undergo a flocking transition in which they acquire a nonzero mean velocity, which corresponds to a spontaneous breaking of the rotation symmetry \cite{Vicsek1995,TonerTu1995}. From experiments and numerical simulations of two dimensional systems, it is now clear that if the landscape becomes too disordered, {\it i.e.}, at high obstacle density, flocking can be inhibited 
and the individuals cannot organize in any ordered fashion \cite{Chepizhko2015,Morin2017,Chardac2021}. At lower obstacle density, on the other hand, quasi-long-range order may emerge  \cite{Toner2018PRL,Chepizhko2015,Morin2017}. The properties of the (quasi) ordered phase somehow depend on the precise nature of the 
interactions  \cite{Ballerini2008,Rahmani2021,Martin2021} or of the disorder \cite{Duan2021,Chardac2021} and on the compressible or incompressible character of the active matter.
Analytical treatments based on the hydrodynamic theory of flocking and the dynamic renormalization group show that long-range order is possible in two and higher dimensions 
for incompressible flocks \cite{chen2022,Chen2022bis}, while it is only possible in dimensions equal to or greater than $3$ for compressible flocks, a quasi-long-ranged phase being then 
predicted in $d=2$  \cite{Toner2018,Toner2018PRL}. The large-scale and long-time scaling behavior is also different in the two cases.

These results provide a rather convincing picture that for active systems whose order parameter exhibits a continuous symmetry, the ordered phase associated with 
collective motion is far more robust to the presence of quenched disorder than it is in their equilibrium counterparts.
All of the above studies are however concerned with the existence and the properties of the ordered or quasi-ordered phase. Here, we consider the flocking 
transition itself, {\it i.e.}, the passage from the disordered phase to the ordered one, which has not yet been investigated in the presence of quenched disorder.

While the study of quenched disorder on active matter is still under development, the equivalent problem at equilibrium has a long history \cite{cardy1996}.
Let us recall some salient properties, focusing for simplicity on the O($N$) model. Excluding spin-glass behavior corresponding to random couplings that are either 
ferromagnetic or anti-ferromagnetic, there are two main ways of introducing disorder in this system. This can be done by modulating the exchange term 
in space ({\it e.g.}, by diluting the material with nonmagnetic impurities) or by adding a magnetic field whose strength varies in space.
The first type of disorder (called random dilution, random mass, or else random temperature) is more benign than the second one and may or may not be a relevant perturbation to the pure system
depending on inequalities on critical exponents known as the Harris criterion \cite{Harris1974,cardy1996}.
The second type of disorder (called random field) is always a relevant 
perturbation.
The upper critical dimension of the transition is shifted from $d=4$ to $d=6$, and the critical behavior of the system is controlled by a so-called zero-temperature fixed
point \cite{cardy1996,Natterman1998}.
Furthermore, perturbation theory to all orders predicts a dimensional-reduction property by which the critical behavior of the random-field 
model is the same as that of the corresponding pure model in two dimensions less \cite{Natterman1998}.
This has been proven wrong in low dimensions and has been shown to be valid only above a critical dimension $d_{dr}(N)<6$ \cite{Tarjus2020}.

A general question that arises is then: to what extent the general picture depicted above for disordered equilibrium systems remains valid for active matter? In this 
article, we address this issue in the case of the flocking transition of active matter at sufficiently high concentrations, for which a description in terms of an incompressible 
active fluid is a good approximation \cite{Toner2015,attanasi2014,Mora2016,Maitra2020,chen2022}.
In this case, the essential of the dynamics is captured by a single vectorial order parameter field, the flow velocity $\bm{v}$ \cite{Toner2015}.
We therefore assume that the dynamics of $\bm{v}$ at long times and long distances is governed by a generalized $d$-dimensional Navier-Stokes equation (NS) augmented 
by a term mimicking the local aligning interactions, a hydrodynamic theory of dry polar active matter known as the Toner-Tu model \cite{TonerTu1995,TonerTu1998}.
To account for the presence of frozen obstacles or heterogeneities in the environment, we further add to the theory a quenched static random force acting on the active fluid. 
The so-obtained model presents a critical point separating an ordered phase characterized by a nonvanishing mean velocity from a disordered phase in which $\bm{v}=0$. 
A renormalization group (RG) analysis shows that at the associated fixed point the aligning interactions vanish: the transition of the active fluid is governed by an 
attractive ``quenched Navier-Stokes'' fixed point. We then show on symmetry grounds that all the other types of quenched disorder that could be considered in place of the 
static random force actually lead to a critical behavior that falls in the same universality class controlled by the quenched Navier-Stokes fixed point. This is in stark contrast 
with the equilibrium case in which, {\it e.g.}, random mass and random field/force correspond to distinct universality classes.

The rest of the paper goes as follows.
In Section~\ref{sec:rf} we present the hydrodynamic model describing an incompressible active fluid in the presence of quenched disorder. We first focus on the effect of a 
quenched, static, random force on the flocking transition and we further comment on the symmetry properties of the theory compared to the thermal case. 
Section~\ref{sec:rep} is devoted to the field-theoretic formulation of the problem and to the replica formalism used to perform the quenched average over the disorder. We derive 
the basic elements of perturbation theory and provide the corresponding one-loop corrections. 
In Section~\ref{sec:RGfunctions},  we obtain the RG functions describing the critical behavior associated with the flocking transition of the system. 
We analyze the fixed points of the resulting RG flow in Section~\ref{sec:critbeh} and we derive the scaling exponents characterizing the new universality class, which we refer  
to as quenched Navier-Stokes.
In Section~\ref{sec:randomMass} we consider the theory under the effect of a random-mass disorder and show that it belongs to the same universality class as the quenched 
random force.
Conclusions and perspectives are provided in Section~\ref{sec:conclusions}.
Finally, in \ref{appendix:details} we detail the computation of the propagator, in \ref{sec_appendixlowd} we discuss the robustness of our perturbative results in lower dimensions, and \ref{sec:symmetries} is devoted to a discussion of the symmetries of the different theories 
considered in the paper.

\section{Disordered incompressible polar active fluid}
\label{sec:rf}
\enlargethispage{\baselineskip}

As established in Refs.~\cite{TonerTu1998,Toner2015}, the hydrodynamic equation governing the evolution of the velocity field $\bm{v}(\bm{x},t)$ in the active fluid reads
\begin{equation}\label{eq:iaf}
  (\tau \partial_t + \lambda\,{v}_\nu \partial_\nu) {v}_\mu = Z \,\partial_\nu\partial_\nu {v}_\mu-\partial_\mu p - \left(a +
  \frac{b}{3!} v^2\right)v_\mu + f_\mu \, ,
\end{equation}
where $\lambda$ is the  advection parameter, $Z$ is a diffusion constant, and the pressure $p$ must adapt to ensure that the incompressibility condition 
\begin{equation}\label{eq:incompressibility}
    \partial_\mu v_\mu  = 0 \,,  \hspace{20pt} \mu = 1,\dots, d \,,
\end{equation}
is fulfilled, where $d$ is the number of space dimensions.  (Hereafter, a sum over repeated indices is always intended.) 
The parameters $a$ and $b$ characterize the local aligning tendency and allow the velocity $\bm{v}$ to take a nonzero magnitude in the uniformly ordered 
phase when $a$ is negative \cite{TonerTu1995}. Stability requires that $b>0$. In Refs.~\cite{TonerTu1998,Toner2015}, the random 
external force $f_\mu(\bm{x},t)$ is assumed to be Gaussian with correlations given by
 \begin{equation}
  \langle f_\mu(\bm{x}_1,t_1) f_\nu(\bm{x}_2,t_2) \rangle = 2D\, \delta_{\mu \nu} \, \delta^{(d)}(\bm{x}_1-\bm{x}_2) \delta(t_1-t_2)\ .
\end{equation}

To model the effect of quenched disorder, we introduce a quenched, i.e., purely static, random force $g_\mu(\bm{x})$ acting on the active fluid:
\begin{equation}
\label{eq:Langevin}
  \partial_t v_\mu = F_\mu[\bm v,p] + f_\mu(\bm{x},t) + g_\mu(\bm{x}) \,,
\end{equation}
where $F_\mu(\bm v,p)$ represents the $t$-local deterministic part of Eq.~\eqref{eq:iaf} and $g_\mu(\bm{x})$ is assumed Gaussian with zero mean and correlations given by
 \begin{equation}
 \label{eq:qnoise}
   \overline{g_\mu(\bm{x}_1) g_\nu(\bm{x}_2)} = \Delta\, \delta_{\mu \nu} \, \delta^{(d)}(\bm{x}_1-\bm{x}_2) \,.
\end{equation}
We will discuss later on additional ways to introduce quenched disorder.

An important distinction between quenched and thermal random forces is worth mentioning.
When there are no aligning interactions [{\it i.e.}, no force $(a + b v^2/6)v_\mu$] and no quenched force, Eq.~\eqref{eq:iaf} is invariant under the 
transformation $\bm{x} \to \bm{x} - \lambda \bm{w} t$ and $\bm{v} \to \bm{v} + \bm{w}$, 
with $\bm{w}$ a generic boost vector.
This holds true in the presence of a thermal noise with temporal correlations given by
$\langle f_\mu(\bm{x}_1,t_1)  f_\nu(\bm{x}_2,t_2) \rangle \propto \delta(t_1-t_2)$ \cite{Medina1989}.
However, such a ``pseudo-Galilean'' symmetry is generally violated in the presence of a quenched disorder.
This can be checked by examining the correlator of a boosted random force $g_\mu(\bm{x} - \lambda \bm{w} t)$ calculated 
from Eq. (\ref{eq:qnoise}), which leads to a statistically inequivalent noise. Generically, quenched disorder induces a preferred reference frame.

\enlargethispage{\baselineskip}
A consequence of the absence of pseudo-Galilean symmetry is that the advection parameter $\lambda$ is no longer protected under renormalization, and therefore 
no exact relations connecting the critical exponents in general dimension $d$ can be deduced \cite{KPZ1986,Medina1989}. This conclusion appears to contrast the results 
of Ref.~\cite{Toner2018}, where  the authors analyze an effective incompressible theory in the presence of the quenched noise and derive exact scaling exponents based 
on the validity of the pseudo-Galilean symmetry.
However, in the latter case one can drop the time coordinate and consider a purely static description 
in which the direction parallel to the mean velocity of the collective motion and the perpendicular directions should be distinguished. One can then show that there 
indeed exists a different sort of pseudo-Galilean symmetry where one replaces the time coordinate by the spatial coordinate in the parallel direction $x_{\parallel}$ (the 
transformation is now $\bm v_\perp \to \bm v_\perp + \bm w_\perp$, $\bm x_\perp \to \bm x_\perp -\lambda \bm w_\perp x_{\parallel}$ for an arbitrary vector 
$\bm w_\perp$ in the perpendicular directions). It is this symmetry that guarantees the nonrenormalizability of the advection parameter $\lambda$.


\section{Replica formalism}
\label{sec:rep}
The stochastic problem associated with Eq. \eqref{eq:Langevin} can be cast in terms of a field theory by using the Martin-Siggia-Rose-Janssen-de Dominicis (MSR-JdD) 
formalism \cite{MSR,Janssen1976,dominicis1976}. For this purpose, we introduce a generating functional for the correlation functions of the velocity field in the presence 
of the quenched disorder,
\begin{equation} 
\begin{aligned}
&    \mathcal Z[\bm g,\bm j]=\\&\big\langle\int \mathcal D \bm v \mathcal D p\, \delta[\partial_t v_\mu - F_\mu[\bm v,p] - f_\mu - g_\mu]
    \delta[\partial_\mu v_\mu] \ee^{\int v_\mu j_\mu}\big\rangle \,,
\end{aligned} 
\end{equation}
where $\delta[\;]$ denotes a delta functional enforced at all points in space and the angular brackets represent an average over the thermal noise. By performing functional 
derivatives of $\mathcal Z$ with respect to $j_\mu$ one generates correlation functions of the velocity field with both the equations of motion and the incompressibility condition 
satisfied. The delta constraints are then expressed in terms of an integral over auxiliary fields $\hat v$ and $\hat p$. The thermal average can be performed explicitly 
and one finally obtains a field theory with twice the number of original fields, $\varphi \equiv \{\hat{\bm v}, \bm v, \hat{p}, p\}$, and action functional 
\begin{equation}
\mathcal S[\bm g,\varphi] =\!\! \int_{\bm{x},t}  \big [ \hat{v}_\mu \left ( \partial_t v_\mu - F_\mu[\bm v,p] - g_\mu(\bm x) \right ) + \hat{p}\,\partial_\mu v_\mu-D \hat{v}_\mu^2  \big ] \,.
\end{equation}
The quenched average over the disorder $\bm{g}$ is greatly simplified by replicating the fields $n$ times, {\it i.e.}, $\varphi \to (\varphi^1, \varphi^2, \dots, \varphi^n)$, and then 
integrating over the corresponding probability distribution function in terms of a cumulant expansion. The process of averaging over the quenched disorder eventually results 
in a coupling between the different replicas that were originally noninteracting. In the presence of external sources 
$\bm J^a \equiv \{ \bm j_{\hat{v}}^a, \bm j^a, j_{\hat{p}}^a, j_{p}^a \}$, with $a=1, \cdots,n$, the replicated partition function reads
\begin{equation}
\begin{aligned}
&  \overline{\prod_{a} \mathcal Z[\bm g,\bm{J}^a]}
 = \overline{ \int ({\prod_a\mathcal D} \varphi^a) ~ \ee^{-\sum_a ~ \mathcal S[\bm g,\varphi^a]+\varphi^a\cdot \bm J^a} } \\
  & = \int (\prod_a \mathcal D \varphi^a) ~ \ee^{-\sum_a S_1[\varphi^a] +\frac{1}{2}\sum_{ab} S_2[\varphi^a,\varphi^b]+\varphi^a\cdot \bm J^a} \,,
\end{aligned}
\end{equation}
where the one-replica contribution $S_1[\varphi^a]$ and the two-replica one $S_2[\varphi^a,\varphi^b]$ are respectively given by
\begin{subequations}\label{eq:replicated-action}
\begin{align}
  S_1 & = \int_{\bm{x},t} \,  \big [\hat{v}^a_\mu \left( \tau \partial_t v^a_\mu -F_\mu[\bm v^a,p^a] \right) + \hat{p}^a\,\partial_\mu v_\mu^a -D( \hat{v}^a_\mu )^2\big ]\,,
  \label{eq:S1}\\
  S_2 & = \Delta  \int_{\bm{x},t_1,t_2} \hat{v}^a_\mu(\bm{x},t_1) \hat{v}^b_\mu(\bm{x},t_2) \,.
  \label{eq:S2}
\end{align}
\end{subequations}
Let us pause for a moment and study the engineering dimension of the coupling constants.
Comparing the $\hat v^2$ terms in the one- and two-replica parts of the action, we immediately see that $\Delta$ has a lower scaling dimension than $D$.
Consequently, we choose to rescale the $\hat v$ field so that the 2-replica action comes with a prefactor 1. This implies that $D$ is nonrenormalizable and must 
not be considered in the perturbative approach. 

The fields $p$ and $\hat p$ only appear in the action in terms that are quadratic. This implies that there are no vertices (and consequently no Feynman diagrams)  with 
$p$ or $\hat p$ leg. Therefore this sector of the theory is not renormalized. The $p$ and $\hat p$ fields are however crucial when computing the 2-point functions. It is convenient to consider the disorder contribution given by $S_2$ as an interaction term.
The propagator of the theory is then given by the inverse of $\delta^2 S_1/\delta \varphi\delta\varphi$ which  is a $(2d+2)\times (2d+2)$ matrix since $\varphi$ represents 2 vector fields ($\bm v$ 
and $\hat {\bm v}$) as well as two scalar fields ($p$ and $\hat p$), see \ref{appendix:details} for some detail. This results in a nontrivial correlation function for $\langle \hat v_\mu v_\nu\rangle$ which reads in Fourier space:\footnote{We adopt the following Fourier convention: $f(\bm{x},t) = \int_{Q} f(\bm{q}, \omega) \ee^{+\ii \omega t - \ii \bm{q} \cdot \bm{x}}$ with shorthand notation
$ Q \equiv (\bm{q}, \omega)$ and
$\int_{Q} \equiv \int {\rm{d}}^d \bm{q} \, {\rm{d}} \omega / (2\pi)^{d+1}$.}
\begin{equation}
   \langle \hat v_\mu  v_\nu\rangle(\bm{q},\omega) = \frac{\PT_{\mu\nu}(\mathbf{q})}{\left(-\ii\,\tau\,\omega + Z\,q^2 + a\right)} \,,
\end{equation}
where $\PT_{\mu\nu} (\bm{q})$ is the transverse projection operator in the momentum representation that is defined as 
$\PT_{\mu\nu} (\bm{q}) = \delta_{\mu\nu} - q_\mu q_\nu/q^2$.
It is convenient to introduce the following diagrammatic representation of the bare propagator and bare interaction vertices:
\begin{subequations}\label{eqs:barequantities}
\begin{align}
\begin{tikzpicture}[baseline=-.1cm]
    \path (0,0) node [below] {$\mu$};
    \draw [-To] (0,0)   -- (1/2,0);
    \draw       (1/2,0) -- (1,0);
    \path (1,0) node [below] {$\nu$};
\end{tikzpicture}
& \equiv \langle \hat v_\mu  v_\nu\rangle(\bm{q},\omega)\,, \label{eq:diagprop}\\
\begin{tikzpicture}[baseline=-.1cm]
    \draw       (0,0)   -- (.4,0);
    \draw [To-] (.2,0)   -- (.4,0);
    \path (0,0) node [above] {$\bm{q}_1+\bm{q}_2\hspace{25pt}$};
    \path (0,0) node [below] {$\mu$};
    \draw  (.4,0)   -- (.8, .4);
    \draw [-To] (.8,.4) -- (.6, .2);
    \path (.8,.4) node [right] {$\nu$};
    \path (.8,.4) node [above] {$\bm{q}_1$};
    \draw  (.4,0)   -- (.8,-.4);
    \draw [-To] (.8,-.4) -- (.6, -.2);
    \path (.8,-.4) node [right] {$\rho$};
    \path (.8,-.4) node [below] {$\bm{q}_2$};
\end{tikzpicture}
& \equiv -\ii \lambda(\delta_{\mu\nu} q_{1,\rho} + \delta_{\mu\rho} q_{2,\nu}) \,, \label{eq:diaglambda}\\
\begin{tikzpicture}[baseline=-.1cm]
    \draw  (0,0)   -- (.4,0);
    \draw [To-] (.2,0)   -- (.4,0);
    \path (0,0) node [below] {$\mu$};
    \draw  (.4,0)   -- (.8, .4);
    \draw [-To] (.8,.4) -- (.6, .2);
    \path (.8,.4) node [right] {$\nu$};
    \draw  (.4,0)   -- (.8,-.4);
    \draw [-To] (.8,-.4) -- (.6, -.2);
    \path (.8,-.4) node [right] {$\rho$};
    \draw  (.4,0)   -- (.8,0);
    \draw [-To] (.8,0) -- (.6,0);
    \path (.8,0) node [right] {$\gamma$};
\end{tikzpicture}
& \equiv \frac{b}{3} \left(\delta_{\mu\nu}\delta_{\gamma\rho} + \delta_{\mu\gamma}\delta_{\nu\rho} + \delta_{\mu\rho}\delta_{\nu\gamma} \right) , \\
\begin{tikzpicture}[baseline=-.1cm]
    \path (0,0) node [above] {$a$};
    \path (0,0) node [below] {$\mu$};
    \draw (0,0)   -- (.4,0);
    \draw [To-] (.2,0)   -- (.4,0);
    \draw [dotted] (.4,0)   -- (.8,0);
    \draw (.8,0) -- (1.2,0);
    \draw [-To] (.8,0)   -- (1,0);
    \path (1.2,0) node [above] {$b$};
    \path (1.2,0) node [below] {$\nu$};
    \filldraw [gray!50] (.4,0) circle (2pt);
    \draw (.4,0) circle (2pt);
    \filldraw [gray!50] (.8,0) circle (2pt);
    \draw (.8,0) circle (2pt);
\end{tikzpicture}
& \equiv \Delta \delta_{\mu\nu}\delta(\omega_1)\delta(\omega_2) \,. \label{eq:diagDelta}
\end{align}
\end{subequations}
We use the standard convention that an ingoing (resp., outgoing) arrow on a vertex represents a $v$ (resp., $\hat v$) field and the direction of the momenta is given by 
the arrows [{\it i.e.}, the momentum $\bm q$ flows from left to right in Eq. \eqref{eq:diagprop} and the momenta $\bm q_1$ and $\bm q_2$ are ingoing in Eq. \eqref{eq:diaglambda}].

Loop corrections to the bare quantities are conveniently computed by introducing the standard effective action $\Gamma[\{\phi^a\}]$ through a Legendre transform of the 
generator $W[\{\bm J^a\}]=\log Z[\{\bm J^a\}]$ of the connected correlation functions  \cite{tauberbook,vasilievbook},
\begin{equation}
    \Gamma[\{\phi^a\}] = -W[\{J^a\}] + \int_{\bm{x},t} \phi^a(\bm{x},t) J^a(\bm{x},t) \,,
\end{equation}
where $\phi^a = \langle \varphi^a \rangle$. The effective action $\Gamma$ can then be computed in terms of the standard 
loop expansion, which at one-loop order is given by
\begin{equation}\label{eq:TrLog}
    \Gamma = S + \frac{1}{2} \Tr \log S^{(2)} \,,
\end{equation}
where $S=\sum_a S_1[\varphi^a] -(1/2)\sum_{ab} S_2[\varphi^a,\varphi^b]$ and $S^{(2)}$ is its second functional derivative. The first term in Eq.~\eqref{eq:TrLog} 
represents the \emph{tree-level} contribution, while the $\Tr\log$ term provides all possible \emph{one-loop} corrections to the effective action. The effective action can 
also be expanded in increasing number of sums over replicas, $\Gamma[\{\phi^a\}]=\sum_a \Gamma_1[\phi^a]-(1/2)\sum_{ab} \Gamma_2[\phi^a,\phi^b]+\cdots$, 
and the renormalizable coupling constants can then be obtained from the appropriate functional derivatives, 
\begin{align}
\Gamma^{(1,1)}_{1,\mu\nu}=
~~\frac{1}{2}~
    \begin{tikzpicture}[baseline=-.1cm,decoration={markings,mark=at position 0.5 with {\arrow{>}}}]
    \coordinate (Vu)     at  (-.4,.4);
    \coordinate (Vd)     at  (-.4,-.4);
    \coordinate (V1)     at  (0,0);
    \coordinate (N1)     at  (1/2,+1/2);
    \coordinate (N2)     at  (1/2,-1/2);
    \draw [dotted] (N1) -- (N2);
    \path (Vu) node [left] {$\mu$};
    \draw[postaction={decorate}]  (V1)   -- (Vu);
    \path (Vd) node [left] {$\nu$};
    \draw[postaction={decorate}]  (Vd)   -- (V1);
    \draw   (V1) arc (180:90:1/2);
    \draw   (V1) arc (180:270:1/2);
    \draw[postaction={decorate}] (N1) arc (90:150:1/2);
    \draw[postaction={decorate}] (N2) arc (270:210:1/2);
    \filldraw [gray!50] (N1) circle (2pt);
    \draw (N1) circle (2pt);
    \filldraw [gray!50] (N2) circle (2pt);
    \draw (N2) circle (2pt);
\end{tikzpicture}
~ - ~
\begin{tikzpicture}[baseline=-.1cm,decoration={markings,mark=at position 0.5 with {\arrow{>}}}]
    \coordinate (Vu)     at  (-.2,.7);
    \coordinate (Vd)     at  (-.2,-.7);
    \coordinate (V1)     at  (0,0);
    \coordinate (V2)     at  (.5-0.35,0.35);
    \coordinate (V3)     at  (.5-0.35,-0.35);
    \coordinate (N1)     at  (1/2,+1/2);
    \coordinate (N2)     at  (1/2,-1/2);
    \draw [dotted] (N1) -- (N2);
    \path (Vu) node [left] {$\mu$};
    \draw[postaction={decorate}]  (V2)   -- (Vu);
    \path (Vd) node [left] {$\nu$};
    \draw[postaction={decorate}]  (Vd)   -- (V3);
    \draw   (V1) arc (180:90:1/2);
    \draw   (V1) arc (180:270:1/2);
    \draw[postaction={decorate}] (N1) arc (90:150:1/2);
    \draw[postaction={decorate}] (N2) arc (270:210:1/2);
    \draw[postaction={decorate}] (V3) arc (225:135:1/2);
    \filldraw [gray!50] (N1) circle (2pt);
    \draw (N1) circle (2pt);
    \filldraw [gray!50] (N2) circle (2pt);
    \draw (N2) circle (2pt);
\end{tikzpicture}
\label{eq:diagGamma111}
~~,
\end{align}

\begin{align}
\Gamma^{(1,2)}_{1,\mu\nu\rho} =
& ~ - ~ 
\begin{tikzpicture}[baseline=-.1cm,decoration={markings,mark=at position 0.5 with {\arrow{>}}}]
    \coordinate (Vu)     at  (-.2,.7);
    \coordinate (Vd)     at  (-.2,-.7);
    \coordinate (V1)     at  (0,0);
    \coordinate (V2)     at  (.5-0.35,0.35);
    \coordinate (V3)     at  (.5-0.35,-0.35);
    \coordinate (V4)     at  (.5-0.35, 0.85);
    \coordinate (N1)     at  (1/2,+1/2);
    \coordinate (N2)     at  (1/2,-1/2);
    \coordinate (A1)     at  (1/2+1/4,1/2);
    \draw [dotted] (N1) -- (N2);
    \path (-0.35,0.35) node [above] {$\nu$};
    \draw[postaction={decorate}]  (-0.35,0.35)   -- (V2);
    \path (Vd) node [left] {$\rho$};
    \draw[postaction={decorate}]  (Vd)   -- (V3);
    \path (V4) node [above] {$\mu$};
    \draw  [-To] (V2)   -- (.5-0.35,0.7);
    \draw        (V2)   -- (.5-0.35,0.85);
    \draw   (V1) arc (180:90:1/2);
    \draw   (V1) arc (180:270:1/2);
    \draw[postaction={decorate}] (N1) arc (90:150:1/2);
    \draw[postaction={decorate}] (N2) arc (270:210:1/2);
    \draw[postaction={decorate}] (V3) arc (225:135:1/2);
    \filldraw [gray!50] (N1) circle (2pt);
    \draw (N1) circle (2pt);
    \filldraw [gray!50] (N2) circle (2pt);
    \draw (N2) circle (2pt);
    \path (A1) node [above] {\color{blue}*};
\end{tikzpicture}
~~~ - ~
\begin{tikzpicture}[baseline=-.1cm,decoration={markings,mark=at position 0.5 with {\arrow{>}}}]
\coordinate (Vu)     at  (-.2,.7);
    \coordinate (Vu)     at  (-.2,-.7);
    \coordinate (Vd)     at  (-.2,+.7);
    \coordinate (V1)     at  (0,0);
    \coordinate (V2)     at  (.5-0.35,-0.35);
    \coordinate (V3)     at  (.5-0.35,0.35);
    \coordinate (V4)     at  (.5-0.35, 0.85);
    \coordinate (N1)     at  (1/2,+1/2);
    \coordinate (N2)     at  (1/2,-1/2);
    \draw [dotted] (N1) -- (N2);
    \path (-0.35,-0.35) node [below] {$\nu$};
    \draw[postaction={decorate}]  (-0.35,-0.35)   -- (V2);
    \path (Vd) node [above] {$\mu$};
    \draw[postaction={decorate}]  (V3)   -- (Vd);
    \path (.5-0.35,-0.85) node [below] {$\rho$};
    \draw  [-To] (.5-0.35,-0.85)   -- (.5-0.35,-0.6);
    \draw        (V2)   -- (.5-0.35,-0.85);
    \draw   (V1) arc (180:90:1/2);
    \draw   (V1) arc (180:270:1/2);
    \draw[postaction={decorate}] (N1) arc (90:150:1/2);
    \draw[postaction={decorate}] (N2) arc (270:210:1/2);
    \draw[postaction={decorate}] (V2) arc (225:135:1/2);
    \filldraw [gray!50] (N1) circle (2pt);
    \draw (N1) circle (2pt);
    \filldraw [gray!50] (N2) circle (2pt);
    \draw (N2) circle (2pt);
\end{tikzpicture}
~~~ + ~ 
\begin{tikzpicture}[baseline=-.1cm,decoration={markings,mark=at position 0.5 with {\arrow{>}}}]
    \coordinate (Vu)     at  (-.2,.7);
    \coordinate (Vd)     at  (-.2,-.7);
    \coordinate (V1)     at  (0,0);
    \coordinate (V2)     at  (.5-0.35,0.35);
    \coordinate (V3)     at  (.5-0.35,-0.35);
    \coordinate (V4)     at  (0.067,0.75);
    \coordinate (V5)     at  (-.4,0);
    \coordinate (N1)     at  (1/2,+1/2);
    \coordinate (N2)     at  (1/2,-1/2);
    \draw [dotted] (N1) -- (N2);
    \path (Vu) node [left] {$\mu$};
    \draw[postaction={decorate}]  (V2)   -- (Vu);
    \path (Vd) node [left] {$\nu$};
    \draw[postaction={decorate}]  (Vd)   -- (V3);
    \path (V5) node [left] {$\rho$};
    \draw[postaction={decorate}]  (V5)   -- (V1);
    \draw   (V1) arc (180:90:1/2);
    \draw   (V1) arc (180:270:1/2);
    \draw[postaction={decorate}] (N1) arc (90:150:1/2);
    \draw[postaction={decorate}] (V1) arc (180:135:1/2);
    \draw[postaction={decorate}] (N2) arc (270:210:1/2);
    \draw[postaction={decorate}] (V3) arc (225:180:1/2);
    \filldraw [gray!50] (N1) circle (2pt);
    \draw (N1) circle (2pt);
    \filldraw [gray!50] (N2) circle (2pt);
    \draw (N2) circle (2pt);
    \path (A1) node [above] {\color{blue}*};
\end{tikzpicture}
\nonumber\\
& ~ + ~
\begin{tikzpicture}[baseline=-.1cm,decoration={markings,mark=at position 0.5 with {\arrow{>}}}]
    \coordinate (Vu)     at  (-.2,.7);
    \coordinate (Vd)     at  (-.2,-.7);
    \coordinate (V1)     at  (0,0);
    \coordinate (V2)     at  (.5-0.35,0.35);
    \coordinate (V3)     at  (.5-0.35,-0.35);
    \coordinate (V4)     at  (0.067,0.75);
    \coordinate (V5)     at  (-.4,0);
    \coordinate (N1)     at  (1/2,+1/2);
    \coordinate (N2)     at  (1/2,-1/2);
    \draw [dotted] (N1) -- (N2);
    \path (Vu) node [left] {$\rho$};
    \draw[postaction={decorate}]  (Vu)   -- (V2);
    \path (Vd) node [left] {$\nu$};
    \draw[postaction={decorate}]  (Vd)   -- (V3);
    \path (V5) node [left] {$\mu$};
    \draw[postaction={decorate}]  (V1)   -- (V5);
    \draw   (V1) arc (180:90:1/2);
    \draw   (V1) arc (180:270:1/2);
    \draw[postaction={decorate}] (N1) arc (90:150:1/2);
    \draw[postaction={decorate}] (V2) arc (135:180:1/2);
    \draw[postaction={decorate}] (N2) arc (270:210:1/2);
    \draw[postaction={decorate}] (V3) arc (225:180:1/2);
    \filldraw [gray!50] (N1) circle (2pt);
    \draw (N1) circle (2pt);
    \filldraw [gray!50] (N2) circle (2pt);
    \draw (N2) circle (2pt);
\end{tikzpicture}
~~,
\end{align}

\begin{align}
\Gamma^{(1,3)}_{1,\mu\nu\rho\sigma} =
& ~ - ~ 
\begin{tikzpicture}[baseline=-.1cm,decoration={markings,mark=at position 0.5 with {\arrow{>}}}]
    \coordinate (Vu)     at  (-.2,.7);
    \coordinate (Vd)     at  (-.2,-.7);
    \coordinate (V1)     at  (0,0);
    \coordinate (V2)     at  (.5-0.35,0.35);
    \coordinate (V3)     at  (.5-0.35,-0.35);
    \coordinate (V4)     at  (.5-0.35,-0.35);
    \coordinate (V5)     at  (.5-0.35,0.85);
    \coordinate (V6)     at  (.5-0.35,-0.85);
    \coordinate (N1)     at  (1/2,+1/2);
    \coordinate (N2)     at  (1/2,-1/2);
    \draw [dotted] (N1) -- (N2);
    \path (-0.35,-0.35) node [below] {$\sigma$};
    \draw[postaction={decorate}]  (-0.35,-0.35)   -- (V4);
    \path (-0.35,0.35) node [above] {$\nu$};
    \draw[postaction={decorate}]  (-0.35,0.35)   -- (V2);
    \path (V6) node [below] {$\rho$};
    \draw[postaction={decorate}]  (V6)   -- (V3);
    \path (V5) node [above] {$\mu$};
    \draw  [-To] (V2)   -- (.5-0.35,0.7);
    \draw        (V2)   -- (.5-0.35,0.85);
    \draw   (V1) arc (180:90:1/2);
    \draw   (V1) arc (180:270:1/2);
    \draw[postaction={decorate}] (N1) arc (90:150:1/2);
    \draw[postaction={decorate}] (N2) arc (270:210:1/2);
    \draw[postaction={decorate}] (V3) arc (225:135:1/2);
    \filldraw [gray!50] (N1) circle (2pt);
    \draw (N1) circle (2pt);
    \filldraw [gray!50] (N2) circle (2pt);
    \draw (N2) circle (2pt);
    \path (A1) node [above] {\color{blue}\quad**};
\end{tikzpicture}
~ - ~ 
\begin{tikzpicture}[baseline=-.1cm,decoration={markings,mark=at position 0.5 with {\arrow{>}}}]
    \coordinate (Vu)     at  (-.2,.7);
    \coordinate (Vd)     at  (-.2,-.7);
    \coordinate (V1)     at  (0,0);
    \coordinate (V2)     at  (.5-0.35,0.35);
    \coordinate (V3)     at  (.5-0.35,-0.35);
    \coordinate (V4)     at  (.5-0.35,-0.35);
    \coordinate (V5)     at  (.5-0.35,0.85);
    \coordinate (V6)     at  (.5-0.35,-0.85);
    \coordinate (N1)     at  (1/2,+1/2);
    \coordinate (N2)     at  (1/2,-1/2);
    \draw [dotted] (N1) -- (N2);
    \coordinate (Vu1)     at  (-.4,.4);
    \coordinate (Vd1)     at  (-.4,-.4);
    \path (Vu1) node [left] {$\mu$};
    \draw[postaction={decorate}]  (V1)   -- (Vu1);
    \path (Vd1) node [left] {$\nu$};
    \draw[postaction={decorate}]  (Vd1)   -- (V1);
    \path (V6) node [below] {$\rho$};
    \draw[postaction={decorate}]  (V6)   -- (V3);
    \path (V5) node [above] {$\sigma$};
    \draw  [-To] (.5-0.35,0.85)   -- (.5-0.35,0.6);
    \draw        (V2)   -- (.5-0.35,0.85);
    \draw   (V1) arc (180:90:1/2);
    \draw   (V1) arc (180:270:1/2);
    \draw[postaction={decorate}] (N1) arc (90:150:1/2);
    \draw[postaction={decorate}] (N1) arc (90:220:1/2);
    \draw[postaction={decorate}] (N2) arc (270:210:1/2);
    \draw[postaction={decorate}] (N2) arc (270:140:1/2);
    \filldraw [gray!50] (N1) circle (2pt);
    \draw (N1) circle (2pt);
    \filldraw [gray!50] (N2) circle (2pt);
    \draw (N2) circle (2pt);
    \path (A1) node [above] {\color{blue}\quad**};
\end{tikzpicture}
\nonumber\\
& ~ + ~
\begin{tikzpicture}[baseline=-.1cm,decoration={markings,mark=at position 0.5 with {\arrow{>}}}]
    \coordinate (Vu)     at  (-.2,.7);
    \coordinate (Vd)     at  (-.2,-.7);
    \coordinate (V1)     at  (0,0);
    \coordinate (V2)     at  (.5-0.35,0.35);
    \coordinate (V3)     at  (.5-0.35,-0.35);
    \coordinate (V4)     at  (.5-0.35,-0.35);
    \coordinate (V5)     at  (.5-0.35,0.85);
    \coordinate (V6)     at  (.5-0.35,-0.85);
    \coordinate (V7)     at  (-.4,0);
    \coordinate (N1)     at  (1/2,+1/2);
    \coordinate (N2)     at  (1/2,-1/2);
    \draw [dotted] (N1) -- (N2);
    \path (-0.35,0.35) node [above] {$\nu$};
    \draw[postaction={decorate}]  (-0.35,0.35)   -- (V2);
    \path (V5) node [above] {$\mu$};
    \draw  [-To] (V2)   -- (.5-0.35,0.7);
    \draw        (V2)   -- (.5-0.35,0.85);
    \path (V5) node [above] {$\mu$};
    \draw  [-To] (V2)   -- (.5-0.35,0.7);
    \draw        (V2)   -- (.5-0.35,0.85);
    \path (V7) node [left] {$\rho$};
    \draw[postaction={decorate}]  (V7)   -- (V1);
    \path (Vd) node [left] {$\sigma$};
    \draw[postaction={decorate}]  (Vd)   -- (V3);
    \draw   (V1) arc (180:90:1/2);
    \draw   (V1) arc (180:270:1/2);
    \draw[postaction={decorate}] (N1) arc (90:150:1/2);
    \draw[postaction={decorate}] (V1) arc (180:135:1/2);
    \draw[postaction={decorate}] (N2) arc (270:210:1/2);
    \draw[postaction={decorate}] (V3) arc (225:180:1/2);
    \filldraw [gray!50] (N1) circle (2pt);
    \draw (N1) circle (2pt);
    \filldraw [gray!50] (N2) circle (2pt);
    \draw (N2) circle (2pt);
    \path (A1) node [above] {\color{blue}\quad\quad*****};
\end{tikzpicture} \,,
\label{eq:diag31}
\end{align}
\begin{align}
\Gamma^{(1,0)(1,0)}_{2,\mu\nu} =\frac 12
\begin{tikzpicture}[baseline=-.1cm,decoration={markings,mark=at position 0.5 with {\arrow{>}}}]
    \coordinate (Vu)     at  (-.2,.7);
    \coordinate (Vd)     at  (-.2,-.7);
    \coordinate (V1)     at  (0,0);
    \coordinate (V1r)     at  (3/2,0);
    \coordinate (V2)     at  (.5-0.35,0.35);
    \coordinate (V3)     at  (.5-0.35,-0.35);
    \coordinate (V4)     at  (0.067,0.75);
    \coordinate (V5)     at  (-.4,0);
    \coordinate (V5r)     at  (1.9,0);
    \coordinate (N1)     at  (1/2,+1/2);
    \coordinate (N2)     at  (1/2,-1/2);
    \coordinate (N3)     at  (1,+1/2);
    \coordinate (N4)     at  (1,-1/2);
    \draw [dotted] (N1) -- (N3);
    \draw [dotted] (N2) -- (N4);
    \path (V5) node [left] {$\mu$};
    \draw[postaction={decorate}]  (V1) -- (V5);
    \path (V5r) node [right] {$\nu$};
    \draw[postaction={decorate}]  (V1r) -- (V5r);
    %
    \draw   (V1) arc (180:90:1/2);
    \draw   (V1) arc (180:270:1/2);
    \draw[postaction={decorate}] (N1) arc (90:180:1/2);
    \draw[postaction={decorate}] (N2) arc (270:180:1/2);
    \draw   (V1r) arc (0:90:1/2);
    \draw   (V1r) arc (0:-90:1/2);
    \draw[postaction={decorate}] (N3) arc (90:0:1/2);
    \draw[postaction={decorate}] (N4) arc (-90:0:1/2);
    \filldraw [gray!50] (N1) circle (2pt);
    \draw (N1) circle (2pt);
    \filldraw [gray!50] (N2) circle (2pt);
    \draw (N2) circle (2pt);
    \filldraw [gray!50] (N3) circle (2pt);
    \draw (N3) circle (2pt);
    \filldraw [gray!50] (N4) circle (2pt);
    \draw (N4) circle (2pt);
\end{tikzpicture}
~~.
\end{align}
The numbers in parenthesis in exponents of $\Gamma$ indicate the number of derivatives with respect to $\hat v$ (first number) and $v$ (second number). In the last equation, 
we need to specify two pairs of number to describe what derivatives are performed for each replica. The blue asterisks count the number of permutations of external $v$ legs 
for each diagram. For instance, the first diagram in Eq. \eqref{eq:diag31} must be summed with the two cyclic permutations of $\{\nu,\rho,\sigma\}$. 
All the  corrections will be evaluated at the upper critical dimension $d_{\rm uc}$.

\section{Renormalization-group functions}
\label{sec:RGfunctions}
From the effective action computed at an RG scale $k$,\footnote{The renormalization-group factor $k$ has the dimension of an inverse length.} we obtain a set of effective coupling constants as
\begin{align}
\Gamma_{1,k}&=\int_{\bm{x},t} \,   \hat{v}^a_\mu \Big[  \tau_k \partial_t v^a_\mu -Z_k \,\partial^2 {v}^a_\mu + \lambda_k\,{v}^a_\nu \partial_\nu {v}^a_\mu +D_k \hat{v}^a_\mu \nonumber\\
&\qquad + a_k v^a_\mu +
  \frac{b_k}{3!} (v^a)^2 v^a_\mu+\partial_\mu p^a \Big] + \hat{p}^a\,\partial_\mu v_\mu^a \,,\label{eq:Gamma1}\\
  \Gamma_{2,k} & = \Delta_k  \int_{\bm{x},t_1,t_2} \hat{v}^a_\mu(\bm{x},t_1) \hat{v}^b_\mu(\bm{x},t_2) \,,\label{eq:Gamma2}
\end{align}
where we have limited ourselves to the terms appearing in the bare action (as will be shown later, all the other coupling constants are easily shown to be nonrenormalizable by simple power counting at the upper critical dimension) and where we have used the fact that the $\{p,\hat p\}$ sector is not renormalized.

To study the critical properties of the system, it is convenient to introduce dimensionless renormalized coupling constants that will be denoted by an index r. To do so, we first 
introduce dimensionless renormalized space and time coordinates, velocity field, and response field,
\begin{subequations}
    \label{eq:dedim}
\begin{align}
    x_\mu&=k^{-1}x_ {\rr,\mu} \,,\\
    t&=\tau_k Z_k^{-1}k^{-2}t_\rr\,,\\
    v_\mu^a&=\Delta_k^{1/2}Z_k^{-1}k^{(d-4)/2}v_{\rr,\mu}^a \,,\\
    \hat v_\mu^a&=\Delta_k^{-1/2}Z_k\tau_k^{-1}k^{(d+4)/2}\hat v_{\rm{r},\mu}^a \,.
\end{align}
\end{subequations}
This rescaling ensures that the first two terms of Eq.~\eqref{eq:Gamma1} and the right-hand side of Eq.~\eqref{eq:Gamma2} appear with a unit prefactor. The dimensionless renormalized coupling 
constants are then easily derived as
\begin{subequations}
\begin{align}
    a_k&=Z_k k^{2}a_{\rr,k} \,,\\
    b_k&=\Delta_k^{-1} Z_k^{3}k^{6-d}b_{\rr,k}\,,\\
    \lambda_k&=\Delta_k^{-1/2}k^{(6-d)/2} Z_k^{2}\lambda_{\rr,k}\,.
\end{align}
\end{subequations}
At the Gaussian fixed point, which is equivalent to considering the linear version of the stochastic evolution equation, the nonlinearities associated with the 
advection term and with the aligning interactions become marginal and the anomalous dimensions associated with $\Delta_k$, $\tau_k$, and $Z_k$ vanish. From the 
above equations it is clear that this happens when $d=6$ which is then the upper critical dimension $d_{uc}$ of the flocking transition. Compared to the thermal case where 
$d_{uc}=4$ \cite{TonerTu1995,TonerTu1998}, the upper critical dimension is thus lifted by $2$ in the presence of a quenched random force \footnote{Note that a different 
scaling applies if one considers an anisotropic situation in which the quenched random force only operates in the directions orthogonal to the flock motion: see Ref.~\cite{Toner2018}.}, as it is for equilibrium systems in the presence of a random field \cite{Natterman1998,Tarjus2020}.

One can check that, as anticipated, the variance $D$ of the thermal noise is indeed nonrenormalizable (and thus corresponds to an irrelevant direction at the critical 
fixed point). By performing the change of variables corresponding to Eq. \eqref{eq:dedim} in the term  $\int_{{\bm x},t } D_k \hat v_\mu^a(\bm{x},t)^2$, we find that 
\begin{equation}
D_k=\Delta_k\tau_k Z_k^{-1}k^{-2}D_{\rr,k} \,.
\end{equation}
At the upper critical dimension $d_{uc}=6$, one finds that $D_k\sim k^{-2}D_{\rr,k}$. The negative power of $k$ implies that the variance $D$ of the thermal noise is 
indeed nonrenormalizable.
In the context of equilibrium random-field models, this amounts to saying that the renormalized temperature is irrelevant and 
that the long-distance behavior is controlled by zero-temperature fixed points \cite{Natterman1998,Tarjus2020}.
It is also easily seen that the coupling constants that are not included in the expressions in Eqs. (\ref{eq:Gamma1}, \ref{eq:Gamma2}) are all nonrenormalizable.

At the fixed point, $Z_k$, $\tau_k$ and $\Delta_k$ behave as power laws of the RG scale $k$ and their associated anomalous dimension are defined according 
to the generic relation
\begin{equation}
    \eta_X=-\frac{\partial \log X_k}{\partial s} \,,
\end{equation}
where $s\equiv \log(\frac{k}{\Lambda})$ is the RG time and $\Lambda$ is an ultra-violet cutoff.
By looking at the scaling of the field, the response field, and the time, we conclude that, at a fixed point (denoted by a star below), these anomalous dimensions are 
related to the critical exponents as follows:
\begin{subequations}
\begin{align}
    z&=2-\eta_Z^*+\eta_\tau^* \,,\\
    \eta&=\eta_Z^* \,,\\
    \bar\eta&=2\eta_Z^*-\eta_\Delta^* \,.
\end{align}
\end{subequations}
From the above results, the response and correlation functions can be put, at large distance and long time, in the  following form

\begin{subequations}
\begin{align}
    \langle \hat v_\mu(\bm 0,0)v_\nu(\bm{x},t)\rangle&\sim\frac {F_1\left(x^{z}/t\right)\delta_{\mu\nu}+F_2\left(x^{z}/t\right)\frac{x_\mu x_\nu}{x^2}}{x^{d-2+\eta+z}} \,,\\
        \langle v_\mu(\bm 0,0)v_\nu(\bm{x},t)\rangle&\sim\frac {F_3\left(x^{z}/t\right)\delta_{\mu\nu}+F_4\left(x^{z}/t\right)
        \frac{x_\mu x_\nu}{x^2}}{x^{d-4+\bar\eta}}\,,
\end{align}
\end{subequations}
where $x=\vert \bm x\vert$ and the $F_i$'s are scaling functions. These functions tend to a constant in the limit of large argument and behave as power laws at small arguments: $F_{1,2}(y)\sim y^{(d-2+\eta+z)/z}$, $F_{3,4}(y)\sim y^{(d-4+\bar\eta)/z}$. The incompressibility condition imposes relations between these functions (not given here).
The presence of the additional anomalous dimension $\bar\eta$ in the long-distance behavior of the 
$2$-point functions is a signature of random-field (or equivalently, random-force) disorder \cite{Natterman1998,Tarjus2020}.

After this general discussion of the renormalization of the theory, we focus on the one-loop results. At this order, several simplifications take place at the level of the $2$-point 
functions. Indeed, when the momentum of the $\hat v$ leg of the 3-point vertex in Eq. \eqref{eq:diaglambda} vanishes, the $v$ legs of this vertex are longitudinal. Since 
these legs must be contracted with a propagator that is transverse, the Feynman diagram vanishes. As a consequence, $\partial_k\tau_k=0$, which implies in particular 
that, at this order, $z=2-\eta$. The diagram renormalizing the two-replica vertex vanishes at zero momentum for the same reason. which implies that, at this order, 
$\bar\eta=2\eta$.

After introducing the rescaled coupling constants $g_1=v_d \lambda_{\rm{r},k}^2$ and $ g_2 =v_d b_{\rm{r},k}$, 
where  the prefactor $v_d^{-1} =2^{d+1}\pi^{d/2}\Gamma(d/2)$, we obtain the following RG beta functions,
\begin{subequations}\label{eqs:RGflow}
\begin{align}
    \beta_{g_1} & = -\varepsilon g_1 + \frac{50}{9} g_1 g_2 + \frac{31}{3} g_1^2 \,,\\
    \beta_{ g_2} & = -\varepsilon g_2 + \frac{73}{6} g_1 g_2 + 15 g_2^2 \,,
\label{eq:m1snu}  
\end{align}
\end{subequations}
together with
\begin{subequations}
\begin{align}
 \left .\frac{\partial \beta_{a_{\rr}}}{\partial a_{\rr}} \right |_{a_\rr=0}& = -2+ \frac{8}{3} g_1 + \frac{80}{9}  g_2  \,,\\
 \eta_Z & = \frac{8}{3} g_1\,, \label{eq:eta}
 \end{align}
\end{subequations}
where $\beta_X=\partial X_{\rr,k}/\partial s$ gives the evolution  of a coupling constant $X_{\rr,k}$ with the RG time $s$. 
(Note that the infrared limit $k\to 0$ is obtained when $s\to -\infty$.)

\section{Critical behavior}
\label{sec:critbeh}

\begin{figure}[t!]
\includegraphics[scale=.6]{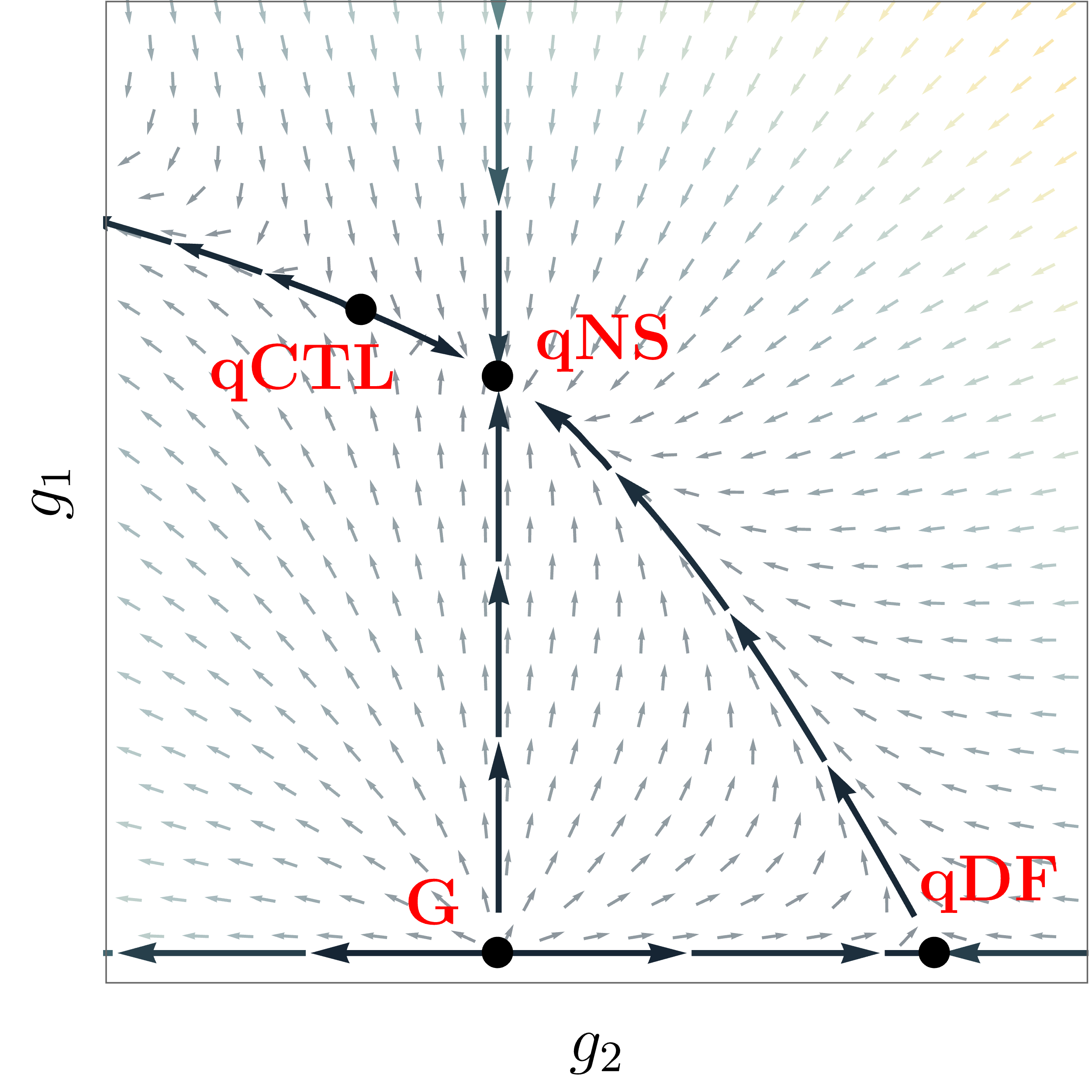}
\caption{One-loop RG flow in the parameter space of the two coupling constants $g_1$ associated with advection and $g_2$ associated with the nonlinear part of the 
aligning interactions. The infrared properties of the theory are governed by the quenched Navier-Stokes (qNS) fixed point. Besides the 
Gaussian fixed point (G), the quenched counterpart of the fixed point found in \cite{Toner2015}  (qCTL), and the quenched dipolar ferromagnet (qDF) fixed point are both unstable. The arrows indicate the 
direction of the RG flow (toward the IR limit) and the shade characterizes the speed of the flow (lightest is fastest). To make the comparison with the same system in the 
absence of a quenched force that has been studied in Ref.~\cite{Toner2015} we put $g_2$ on the horizontal axis and $g_1$ on the vertical one.}
\label{Fig:RGflow}
\end{figure}

The RG equations in Eqs. \eqref{eqs:RGflow} admit four fixed points whose location in the $(g_2,g_1)$ diagram is illustrated in Fig.~\ref{Fig:RGflow}.
The infrared properties of the system are governed by a stable advective fixed point ($g_1^*=3\varepsilon/31, g_2^*=0$), which represents a new 
``quenched Navier-Stokes'' (qNS) universality class. This fixed point is reminiscent of the one describing the large-distance and long-time properties of a stirred fluid 
described by the Navier-Stokes equation that is forced at zero wavelength \cite{Forster1977}. However, the forcing here is due to a quenched, static, random force, which 
produces a distinct universality class.
In addition to  the Gaussian (G) fixed point $(g_1^*=g_2^*=0)$, there are two other unstable fixed points. One corresponds to a fixed point with no advection ($g_1^*=0,g_2^*=\varepsilon/15$) and describes the critical behavior of  an isotropic ferromagnet with long-ranged dipolar interactions in the 
presence of quenched disorder (qDF). The corresponding pure case was studied in \cite{Fisher1973}.
The other unstable fixed point $(g_1^*=51\varepsilon/472, g_2^*=-99\varepsilon/4720)$ corresponds to a theory with both 
advection and interactions (qCTL).
This fixed point is the quenched counterpart 
of the attractive fixed point found in an incompressible active fluid in  
the presence of thermal noise only~\cite{Toner2015}. The effect of the quenched disorder is to drive the fixed point to the nonphysical region where $g_2<0$ and 
to change its stability.

Most critical exponents associated with the new qNS universality class can be determined in the usual way.
The anomalous dimension is obtained by evaluating Eq.~\eqref{eq:eta} at the fixed point, which yields 
\begin{equation}
    \eta=\frac 8{31}\varepsilon \,,
\end{equation}
from which we obtain that
\begin{align}
    \bar\eta & = \frac{16}{31}\varepsilon \,,\\
    z & =2-\frac 8{31}\varepsilon \,.
\end{align}
Eq.~\eqref{eq:m1snu} evaluated at the fixed point characterizes the speed at which the relevant operator (the ``mass term", associated with the departure from 
the critical temperature) increases. This leads to the determination of the critical exponent
\begin{equation}
   \nu=\frac 12+\frac 2{31} \varepsilon \,,
\end{equation}
associated with the growth of the correlation length.

On the other hand, the determination of the exponent $\beta$ that characterizes the increase of the order parameter is more subtle here. Indeed, the ordered phase 
appearing at the critical point is one in which there is a nonzero mean velocity point \cite{chen2022}: the critical point corresponds to a flocking transition to 
coherent motion. Surprisingly, however, at the associated qNS fixed point the renormalized aligning interactions vanish. In particular, the corresponding nonlinearity 
(the $v^4$ coupling constant in the potential from which the local aligning force derives) $g_2^*=0$, and it is not possible to determine the average value of the velocity field 
right at the fixed point. In such a case, one cannot simply make use of the conventional hyperscaling relation, $\beta=\nu(d-2+\eta)/2$.

An operator such as the $v^4$ coupling constant ($g_2$) at the qNS fixed point is called  dangerously irrelevant.\footnote{The simplest occurrence of a dangerously 
irrelevant is in the Ising model in $d>4$. In this situation, the Gaussian fixed point is stable. The $\phi^4$ coupling constant is driven to zero, which induces a violation of 
the hyperscaling relation, which eventually leads to the critical exponent $\beta=1/2$, independently of the dimension (as long as $d>4$.)} In the presence of such an 
operator, the hyperscaling function is modified to
\begin{equation}
    \beta=\nu(d-2+\eta-\omega)/2,
\end{equation}
where $\omega$ characterizes the speed at which the $v^4$ coupling term vanishes, $\omega=\frac{\partial \beta_{g_2}}{\partial g_2}|_{g_1^*,g_2^*}$. We then find that 
\begin{equation}
\beta=\frac 12- \frac{41}{248}\varepsilon \,,
\end{equation}
which completes the characterization of the critical behavior at one-loop order.
In \ref{sec_appendixlowd}, through approximations that are less controlled than the present $\varepsilon$-expansion, we show that the irrelevance of the nonlinearity associated with the aligning interactions at the qNS fixed point persists in lower dimensions.

Note that the critical scaling of the flocking transition does not satisfy the property of dimensional reduction that is found in random-field models at 
equilibrium and even in the out-of-equilibrium quasi-statically driven random-field Ising model when studied in the vicinity of the upper critical dimension. As previously 
stressed, the fixed point that controls the transition in the presence of a quenched random force/field is of different nature than the fixed point in the absence of 
quenched disorder. However, the remark also applies to the other non-Gaussian fixed points, in particular to the quenched dipolar-ferromagnet one, whose critical 
exponents at order $\varepsilon$ do not coincide with those of the pure system, despite the absence of the advection term.\footnote{In the present case, there is  no direct analog of the Parisi-Sourlas supersymmetry in the random-field Ising model \cite{Parisi1979}.
This is due to the fact that the field is a  $d$-dimensional vector in Euclidean space whose upgrade as a superfield in superspace is not as straightforward as for the scalar field describing magnetization.}

To conclude this section, we study the fate of the pseudo-Galilean invariance at the qNS fixed point. As already stated, the quenched disorder explicitly breaks the 
pseudo-Galilean invariance at the microscopic level, but the symmetry could nonetheless emerge at large scale in the renormalized theory. For this to happen, it would be 
necessary that the advection term $\lambda_k$ flows during a transient regime and then tends to a constant. At the qNS fixed point, the renormalized coupling constant 
$\lambda_{k,\rr}$ takes a nonzero value. In consequence, the advection term behaves as $\lambda_k\sim k^{\epsilon/2-2\eta}\sim k^{-\epsilon/62}$ in the scaling regime 
and does not go to a finite limit. There is thus no restoration of a pseudo-Galilean symmetry at criticality.

\section{A unique universality class in the presence of quenched disorder}
\label{sec:randomMass}

The aim of this  section is to investigate whether there exists other universality classes for the flocking transition of disordered polar active fluids on top of the quenched 
Navier-Stokes one that we have just found. To put this issue in context, let us briefly recall the situation of the Ising model at equilibrium. 
In this case, the disorder can couple either to the local magnetization $\phi(\bm x)$ (random field), with a term of the form $-\int _x h(\bm x)\phi(\bm x)$, or to the energy 
density (random mass, random temperature or random dilution), with a term of the form $\int _{\bm x} \delta a(\bm x)\phi(\bm x)^2$. These two types of disorder belong to 
different universality classes because their symmetry properties are different: the random field breaks the $\mathbb Z_2$ symmetry in each sample while the random 
mass preserves it. These different symmetry properties can be seen at the level of the $2$-replica action which is proportional to $\int_{\bm x}\phi_a(\bm x)\phi_b(\bm x)$ 
for the former and to $\int_{\bm x}\phi_a(\bm x)^2\phi_b(\bm x)^2$ for the latter. We observe that the random-mass action is invariant if we flip one replica field, 
$\phi^a\to-\phi^a$, while the random-field action is only invariant if all replica fields are flipped simultaneously. 

Is the situation similar in the context of polar active matter? To address this question, we consider the case where a quenched random fluctuation $\delta a(\bm x)$ 
is added to the mass $a$ in Eq.  \eqref{eq:iaf}, such that
\begin{align}
     \overline{\delta a(\bm{x})} = 0, \ \ \overline{\delta a(\bm{x}_1) \delta a(\bm{x}_2)} = \Delta_a \ \delta^{(d)}(\bm{x}_1-\bm{x}_2).
\end{align}
The dynamical equation then becomes
\begin{equation}
\label{eq:iaf_QM}
\partial_t v_{\mu} = F_{\mu}[\bm v,p] + f_{\mu} + \delta a(\bm x) \ v_{\mu} \,,
\end{equation}
where as before $F_{\mu}[\bm v,p]$ represents the deterministic part of Eq. (\ref{eq:iaf}). 
We can repeat the MSR-JdD construction described in Sect.~\ref{sec:rep}. The expression of the 1-replica action remains unchanged, see Eq.\eqref{eq:S1}, while 
the 2-replica action now takes the form
\begin{align}
  S_2 & = \Delta_a  \int_{\bm{x},t_1,t_2} \hat{v}^a_\mu(\bm{x},t_1) v^a_\mu(\bm{x},t_1) \hat{v}^b_\nu(\bm{x},t_2) v^b_\nu(\bm{x},t_2) \,.
  \label{eq:S2_QM}
\end{align}
At first sight, one may think that the symmetry group of this theory is larger than in the situation treated in Section~\ref{sec:rf} because the vector indices of identical replicas 
are contracted together while in Eq.~\eqref{eq:S2_QM} the vector indices of different replicas are contracted. This conclusion, however, turns out to be wrong because the 
vector indices are not mere internal labels but are coupled to space derivatives through the advection term. The transformation that leaves the action invariant involves both 
a rotation of the vector indices {\em and} the point at which the field is evaluated.

This implies in particular that the 2-replica action of the random mass problem is not invariant under a rotation of one replica only. It is nonetheless invariant if we rotate all 
the replica fields simultaneously, as is Eq.~\eqref{eq:S2}. We conclude that there is no symmetry difference between the two types of disorders which, consequently, belong 
to the same universality class whenever there is advection. More generally, it is not possible to engineer a disorder which would yield more symmetry than that experienced 
by the random field. In particular this also applies to a random anisotropy term that would correspond to adding $\delta a_{\mu\nu} v_\nu$ in 
Eq. (\ref{eq:iaf_QM}) and 
$\Delta_a  \int_{\bm{x},t_1,t_2} \hat{v}^a_\mu(\bm{x},t_1) v^a_\nu(\bm{x},t_1)[ \hat{v}^b_\mu(\bm{x},t_2) v^b_\nu(\bm{x},t_2)$ $+ \hat{v}^b_\nu(\bm{x},t_2) v^b_\mu(\bm{x},t_2)]$ in Eq. (\ref{eq:S2_QM}). We conclude that there is just one universality class for the transition in polar active matter in the presence of quenched disorder, not several as in the equilibrium 
case. (We leave apart here the possibility of introducing long-range correlations of the disorder, for instance through a quenched random pressure gradient which may lead to an effective random gauge field \cite{Chardac2021}, as well as  spin-glass-like behavior associated with random aligning and anti-aligning 
forces.)

Another way of checking that random mass and random field leads to the same universality class is by means of Feynman diagrams.
We represent the vertices associated with the thermal diffusion and the random mass in Eq. \eqref{eq:S2_QM}  as
\begin{align}
\begin{tikzpicture}[baseline=-.1cm,decoration={markings,mark=at position 0.5 with {\arrow{>}}}]
    \coordinate (Vl)     at  (0,0);
    \coordinate (Vr)     at  (1.2,0);
    \coordinate (V1)     at  (0.6,0);
    \path (Vl) node [below] {$\mu$};
    \path (Vr) node [below] {$\nu$};
    \draw[postaction={decorate}]  (V1)   -- (Vl);
    \draw[postaction={decorate}]  (V1)   -- (Vr);
    \filldraw [black] (.6,0) circle (2pt);
\end{tikzpicture}
& \equiv 2D \delta_{\mu\nu} \,,
\end{align}
\begin{align}
& \begin{tikzpicture}[baseline=-.1cm,decoration={markings,mark=at position 0.5 with {\arrow{>}}}]
    \coordinate (Vu)     at  (-.5,.4);
    \coordinate (Vd)     at  (-.5,-.4);
    \coordinate (N1)     at  (0,0);
    \coordinate (N2)     at  (1/2,0);
    \coordinate (Wu)     at  (1,.4);
    \coordinate (Wd)     at  (1,-.4);
    \draw [dotted] (N1) -- (N2);
    \draw[postaction={decorate}]  (N1)   -- (Vu);
    \draw[postaction={decorate}]  (Vd)   -- (N1);
    \draw[postaction={decorate}]  (N2)   -- (Wu);
    \draw[postaction={decorate}]  (Wd)   -- (N2);
    \draw[black,fill=gray!50] (N1) circle (2pt);
    \draw[black,fill=gray!50] (N2) circle (2pt);
    \path (Vu) node [above] {$\bm{q}_1,\omega_1$};
    \path (Wu) node [above] {$\bm{q}_4,\omega_4$};
    \path (Vd) node [below] {$\bm{q}_2,\omega_2$};
    \path (Wd) node [below] {$\bm{q}_3,\omega_3$};\,.
\end{tikzpicture} \,. \label{eq:qrm_vertex}
%
\end{align}
One can find a 3-loop diagram that generates the random-field vertex \eqref{eq:diagDelta} out of the random-mass vertex:
\begin{equation}
\begin{tikzpicture}[baseline=-.1cm,decoration={markings,mark=at position 0.5 with {\arrow{>}}}]
    \coordinate (VLu)     at  (-.5,.75);
    \coordinate (VLd)     at  (-.5,-.75);
    \coordinate (N1)     at  (0,0);
    \coordinate (N2)     at  (.75,.75);
    \coordinate (VCuu)     at  (0,.75);
    \coordinate (VCdd)     at  (0,-.75);
    \coordinate (VCu)     at  (0,.25);
    \coordinate (VCd)     at  (0,-.25);
    \coordinate (VRu)     at  (1,.25);
    \coordinate (VRd)     at  (1,-.25);
    \coordinate (V2u)     at  (0.5,.75);
    \coordinate (V2d)     at  (0.5,-.75);
    \coordinate (Wu)     at  (.75,1);
    \coordinate (Wd)     at  (.75,-1);
    \coordinate (W2u)     at  (1,.75);
    \coordinate (W2d)     at  (1,-.75);
    \draw [dotted] (VCuu) -- (VCdd);
    \draw [dotted] (VRu) -- (VRd);
    \draw[postaction={decorate}]  (VCuu)   -- (VLu);
    \draw[postaction={decorate}]  (VCdd)   -- (VLd);
    \draw[postaction={decorate}]  (V2u)   -- (VCuu);
    \draw[postaction={decorate}]  (V2d)   -- (VCdd);
    \draw[postaction={decorate}]  (W2u)   -- (VRu);
    \draw[postaction={decorate}]  (W2d)   -- (VRd);
    \draw[postaction={decorate}]  (VRu)   -- (V2u);
    \draw[postaction={decorate}]  (VRd)   -- (V2d);
    \draw[postaction={decorate}]  (W2u)   -- (V2u);
    \draw[postaction={decorate}]  (W2d)   -- (V2d);
    \draw[black,fill=gray!50] (VCuu) circle (2pt);
    \draw[black,fill=gray!50] (VCdd) circle (2pt);
    \draw[black,fill=gray!50] (VRu) circle (2pt);
    \draw[black,fill=gray!50] (VRd) circle (2pt);
    \draw[black,fill=black] (W2u) circle (2pt);
    \draw[black,fill=black] (W2d) circle (2pt);
\end{tikzpicture} ~~~.
\end{equation}
The presence of this diagram implies that, starting from the random-mass only case, a random field will be generated along the RG flow. 
We also note that this diagram involves the advection term, which is responsible for breaking the symmetry down to $O_{\text{diag.}}(d)$

\section{Conclusion}
\label{sec:conclusions}
In this work, we have investigated how quenched disorder affects the critical scaling of the flocking transition in a polar incompressible active fluid. We show 
that the transition belongs to a new universality class, which is the same for all forms of static disorder. At one-loop order in a dimensional expansion around the 
upper critical dimension $d_{uc}=6$, we find that the critical properties of the disordered system are described by a quenched Navier-Stokes fixed point at which the aligning interactions are irrelevant but dangerously so. The nature of this fixed point is thus very different from its counterpart in the pure (thermal) case \cite{Toner2018},  
suggesting that the flocking transition in polar active matter is very sensitive to quenched perturbations.

To access the critical properties of the disordered active fluid, we have presented for the first time a renormalization group (RG) analysis of an active system based on a  replica formalism for handling the cumulants of the renormalized quenched disorder.
Our analysis could be extended to higher orders of the perturbative RG or to a nonperturbative functional implementation of the RG \cite{nprg-rev} (to check in particular if the quenched disorder does not lead to a nonanalytic fixed-point theory in low dimensions, as it does in equilibrium systems in the presence of a random field \cite{Tarjus2020}).
It can also be generalized to characterize the dynamical 
scaling properties of active matter in other instances where the system evolves in a static heterogeneous environment: to stay within the realm of polar active matter, one 
could consider the case of a compressible fluid whose (quasi) ordered phase has been recently studied \cite{Toner2018}, or add the effect of inertia 
\cite{Cavagna2021} or of spatial anisotropy \cite{solon2022}.

While special attention has been paid to the polar alignment of active matter, little is known about the influence of disorder on the collective properties of 
scalar active matter, for which the order parameter is the density field. Recently, disorder was shown to suppress the so-called motility-induced phase separation in $d=2$ in favor 
of an asymptotically homogeneous phase with anomalous long-ranged correlations \cite{Kardar2021}. Further studies could therefore investigate the role of quenched 
disorder in relation with the recently discovered bubbly phase \cite{Elsen2018,Xia-qing2020} or in the case of the collective properties of assemblies of chemotactic 
particles \cite{Ric2021,Ric2022}.
\\

\section*{Aknowledgments}
We thank Hugues Chat\' e and Alexandre Solon for fruitful discussions.
RBAZ acknowledges the support from the French ANR through the project NeqFluids (grant ANR-18-CE92-0019).

\appendix
\section{Calculation of the propagator}
\label{appendix:details}

As explained in the main text, we consider the propagator obtained by inverting the 2-point function but we leave aside the $\hat v^2$ terms appearing in the 
$2$-replica action $S_2$ because we treat them as interactions. We also recall that $D$ is not renormalizable and can be discarded in a perturbative calculation. We 
therefore have to invert the $(2d+2)\times (2d+2)$ matrix
\begin{align*}
 \mathbb{A} =   \begin{pNiceMatrix}[first-row,first-col]
   & \hat v_\nu & v_\nu & \hat p & p       \\
\hat v_\mu & 0   & X\delta_{\mu\nu}   & 0   & -\ii q_\mu    \\
v_\mu & X^\star\delta_{\mu\nu}   & 0   & \ii q_\mu   & 0    \\
\hat p & 0   & -\ii q_\nu   & 0   & 0   \\
 p & \ii q_\nu   & 0   & 0   & 0  
\end{pNiceMatrix} \,,
\end{align*}
where we have used the shorthand notation $X=\ii \tau \omega+Zq^2+a$. The inverse of this matrix is easily found to be
\begin{align*}
\mathbb{A}^{-1}=   \begin{pNiceMatrix}[first-row,first-col]
 & \hat v_\rho & v_\rho & \hat p & p       \\
\hat v_\nu & 0   & \PT_{\nu\rho}/{X^\star}   & 0   & -\ii {q_\nu}/{q^2}    \\
v_\nu & {\PT_{\nu\rho}}/{X}   & 0   & \ii{q_\nu}/{q^2}    & 0    \\
\hat p & 0   & -\ii {q_\rho}/{q^2}   & 0   & X^\star/q^2   \\
 p & \ii {q_\rho}/{q^2}  & 0   & X/q^2   & 0  
\end{pNiceMatrix} \,.
\end{align*}
Not all the propagators involving the fields $p$ and $\hat p$ vanish. They are however not needed in actual calculations of the Feynman diagrams because there are no vertex involving these fields. The only propagator that appears in the calculation of the Feynman diagrams is
\begin{align*}
    \langle \hat v_\mu v_\nu\rangle(Q)=\frac{\PT_{\mu\nu}}{-\ii\tau \omega+Z q^2+a} \,.
\end{align*}
The fields $p$ and $\hat p$ are responsible for the fact that this propagator is purely transverse.
\\
\section{Stability of the qNS fixed point as a function of dimension}
\label{sec_appendixlowd}
 In this appendix, we describe preliminary studies of the critical behavior of disordered polar active matter in dimensions below $d_{uc}=6$ beyond the $\varepsilon$ expansion. We have considered two generalizations of the calculation presented in the main text: the fixed-dimension perturbative renormalization group and the nonperturbative renormalization group (NPRG) (see Ref. \cite{nprg-rev} for a review). 
 
 The NPRG is a modern implementation of Wilson's original ideas. It consists in computing an effective average action $\Gamma_k$ obtained by freezing the fluctuations of momenta smaller than the RG scale $k$ through the addition of a term quadratic in the fields. The evolution of $\Gamma_k$ when changing $k$ is governed by an exact functional equation (the Wetterich equation \cite{Wetterich1993}), which has proven to be a good starting point for devising approximate RG flows \cite{nprg-rev}. In a nutshell, the idea is to consider a truncation of $\Gamma_k$, characterized by a finite number of coupling constants. At each step of the RG flow, one projects the exact flow on this truncation and one thus obtains a set of coupled differential equations. Of course, the richer the truncation, the more reliable the results. 
 
 In this section, we discuss the simplest NPRG implementation of Eqs.~(\ref{eq:Gamma1},\ref{eq:Gamma2}), with $D_k=0$.\footnote{We use the so-called optimized regulator \cite{nprg-rev} and we neglect the dependence of the resulting threshold functions on the anomalous dimension for simplicity. We have checked that adding it does not change the qualitative picture.} The beta functions obtained through this procedure read
 \begin{align}
\label{eq:NPRG1}  \beta_a&= -2 a -\frac{8 a \left(-3 d^2+d+6\right) {g_1}}{(a+1)^4 d^2 (d+2)}-\frac{8 (d-1) (d+2) {g_2}}{3 (a+1)^3 d^2}\\
  \beta_{g_1}&=(d-6) g_1+\frac{16 (d (9 d-1)-18) {g_1} {g_2}}{3 (a+1)^4 d^2 (d+2)}\nonumber\\
  &+\frac{32 {g_1}^2 ((a+1) (d+2) (d (3 d-1)-6)-4 d)}{(a+1)^5 d^2 (d+2)^2}\\
  \beta_{g_2}&=(d-6) {g_2}+\frac{8 (d-1) (d (d+10)+12) {g_2}^2}{(a+1)^4 d^2 (d+2)}\nonumber\\
  &+g_1g_2\left(\frac{24 \left(3 d^2-d-6\right)}{(a+1)^4 d^2 (d+2)}+\frac{32 (d-1) (d+4)}{(a+1)^5 d (d+2)^2}\right)
 \end{align}
 and the anomalous dimension is
 \begin{equation}
\label{eq:NPRG2} \eta =\frac{8 \left(3 d^2-d-6\right) {g_1}}{(a+1)^4 d^2 (d+2)}\,.
\end{equation}
 We retrieve the 1-loop results quoted in the main text by evaluating the beta functions for $d=6-\epsilon$ and expanding them around $a=0$. Indeed, at the perturbative fixed points, $g_1\sim g_2\sim\epsilon$ and Eq.~(\ref{eq:NPRG1}) implies that $a\sim\epsilon$.
 
\begin{figure}
     \centering
     \includegraphics[width=\linewidth]{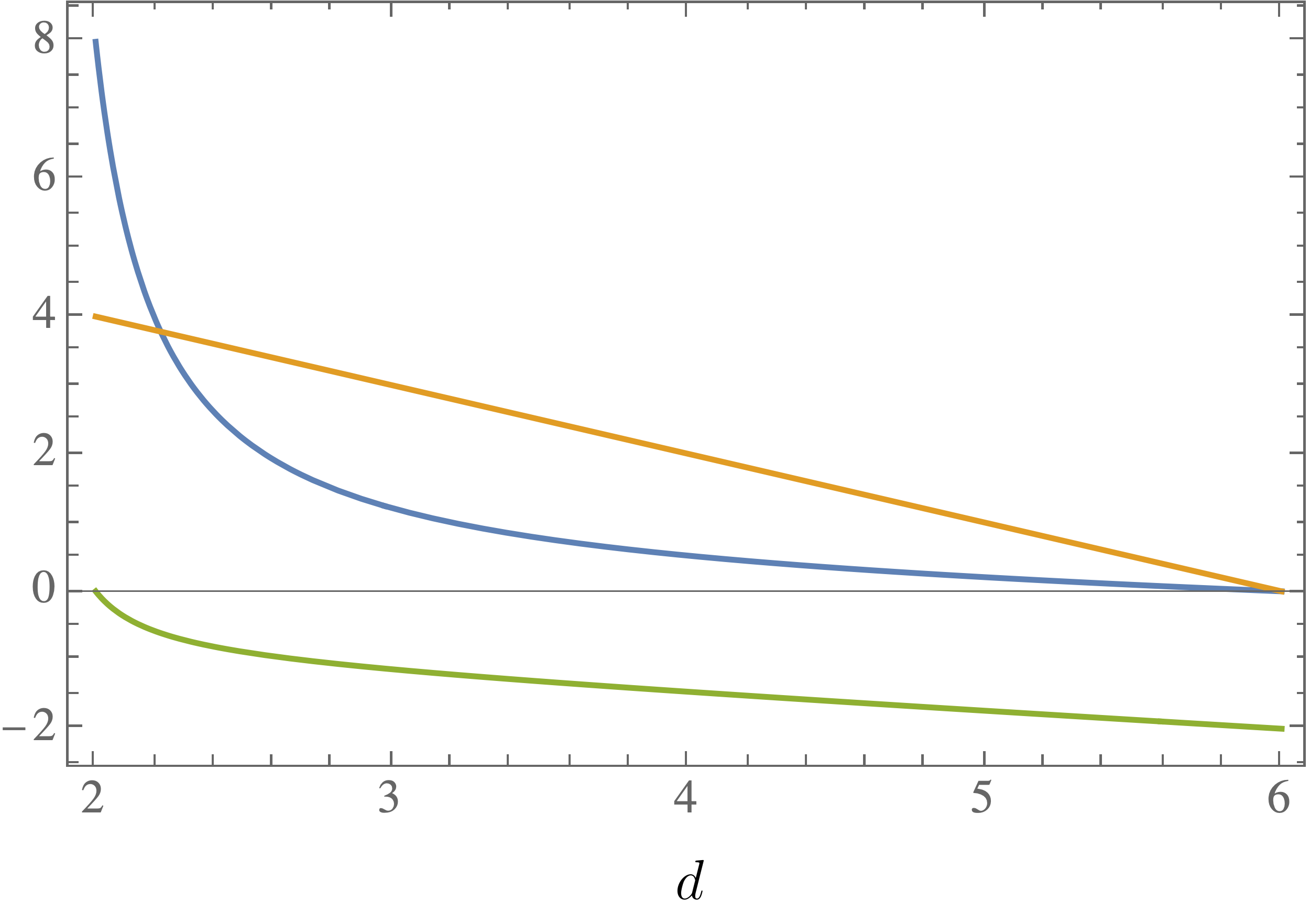}
     \caption{Eigenvalues of the stability matrix around the qNS fixed point in the NPRG calculation. The green line represents the unstable direction associated with $a$, the orange is the stable direction associated with $g_1$, which is equal to $6-d$ in this approximation, and the blue line is the nontrivial stability exponent of the $g_2$ direction. We find that it remains positive for all dimensions.}
     \label{fig_NPRG}
 \end{figure}

 Focusing now on the qNS fixed point ($a=g_2=0$), the solution for $g_1$ at this fixed point is easily determined as a function of $d$. We can then study the stability matrix, which admits one negative eigenvalue (unstable direction) and two stable ones, which both start linearly around $d=6$, see Fig.~\ref{fig_NPRG}. The eigenvalue associated with the $g_2$ direction shows no sign of decreasing. Based on these preliminary results, it seems that the qNS fixed point governs the critical physics in all dimensions smaller than 6. 
 We observe that the negative eigenvalue tends to zero when $d\to 2$, which would indicate a divergence of the critical exponent $\nu$. The validity of our approximation in this regime is however questionable because we also find that the anomalous dimension tends to 2 in this limit, which is quite large.  
 
In addition, we have checked that this general picture also holds true in a fixed-dimension perturbative RG computation at one loop.(In this calculation all the corrections coming from the one-loop diagrams at computed for a generic dimension $d$ instead of $d=6$ as in the main text.)

\section{Symmetries in disordered polar active matter}
\label{sec:symmetries}

This appendix is devoted to the symmetries of the different MSR-JdD actions considered in this article. To start with, we consider the pure system in the absence of advection [$\lambda=0$ in Eq.~\eqref{eq:S1}] and without the fields $p$ and $\hat p$. The theory is invariant under the rotation group $O(d)$ whose infinitesimal variation, 
which is characterized by an antisymmetric $d\times d$ matrix $\omega$ ($\omega_{\mu\nu}=-\omega_{\nu\mu}$), transforms the vector indices as 
$\delta v_\mu=\omega_{\mu\nu}v_\nu$ and a similar transformation for the $\hat v$ field. The theory is also invariant under another rotation group $O(d)$ which acts on 
the point $\bm x$ at which the field is evaluated: $\delta v_\mu=\omega_{\alpha\beta}x_\alpha\partial_\beta v_\mu$ and similarly for $\hat v$. In the presence of the advection 
term, the symmetry $O(d)\times O(d)$ is broken down to a $O_{\text{diag}}(d)$ group whose infinitesimal transformation reads: $\delta v_\mu=\omega_{\mu\nu}v_\nu+\omega_{\alpha\beta}x_\alpha\partial_\beta v_\mu$. In the language of high-energy physics, the first piece of the transformation would be called the spin part and the 
second one the orbital part. The pure theory is invariant under $O_{\text{diag}}(d)$, which is broken by a random field, a random mass,
and a random anisotropy. There is no way to 
differentiate these forms of quenched disorder in terms of symmetry, and we conclude that there is just one universality class of disordered polar active systems 
(provided advection does not vanish at the fixed point).

\bibliographystyle{biblio}
\bibliography{biblio.bib}

\providecommand{\href}[2]{#2}\begingroup\raggedright\begin{thebibliography}{10}

\bibitem{Vicsek2012review}
T.~Vicsek and A.~Zafeiris, \emph{Collective motion},
  \href{https://doi.org/https://doi.org/10.1016/j.physrep.2012.03.004}{\emph{Physics
  Reports} {\bfseries 517} (2012) 71}.

\bibitem{Chepizhko2013}
O.~Chepizhko, E.G.~Altmann and F.~Peruani, \emph{Optimal noise maximizes
  collective motion in heterogeneous media},
  \href{https://doi.org/10.1103/PhysRevLett.110.238101}{\emph{Phys. Rev. Lett.}
  {\bfseries 110} (2013) 238101}.

\bibitem{Peruani2018}
F.~Peruani and I.S.~Aranson, \emph{Cold active motion: How time-independent
  disorder affects the motion of self-propelled agents},
  \href{https://doi.org/10.1103/PhysRevLett.120.238101}{\emph{Phys. Rev. Lett.}
  {\bfseries 120} (2018) 238101}.

\bibitem{Das2018}
R.~Das, M.~Kumar and S.~Mishra, \emph{Polar flock in the presence of random
  quenched rotators},
  \href{https://doi.org/10.1103/PhysRevE.98.060602}{\emph{Phys. Rev. E}
  {\bfseries 98} (2018) 060602}.

\bibitem{Sandor2017}
C.~S\'andor, A.~Lib\'al, C.~Reichhardt and C.J.~Olson~Reichhardt, \emph{Dynamic
  phases of active matter systems with quenched disorder},
  \href{https://doi.org/10.1103/PhysRevE.95.032606}{\emph{Phys. Rev. E}
  {\bfseries 95} (2017) 032606}.

\bibitem{Vicsek1995}
T.~Vicsek, A.~Czir\'ok, E.~Ben-Jacob, I.~Cohen and O.~Shochet, \emph{Novel type
  of phase transition in a system of self-driven particles},
  \href{https://doi.org/10.1103/PhysRevLett.75.1226}{\emph{Phys. Rev. Lett.}
  {\bfseries 75} (1995) 1226}.

\bibitem{TonerTu1995}
J.~Toner and Y.~Tu, \emph{Long-range order in a two-dimensional dynamical
  $\mathrm{XY}$ model: How birds fly together},
  \href{https://doi.org/10.1103/PhysRevLett.75.4326}{\emph{Phys. Rev. Lett.}
  {\bfseries 75} (1995) 4326}.

\bibitem{Chepizhko2015}
O.~Chepizhko and F.~Peruani, \emph{Active particles in heterogeneous media
  display new physics},
  \href{https://doi.org/10.1140/epjst/e2015-02460-5}{\emph{The European
  Physical Journal Special Topics} {\bfseries 224} (2015) 1287}.

\bibitem{Morin2017}
A.~Morin, N.~Desreumaux, J.-B.~Caussin and D.~Bartolo, \emph{Distortion and
  destruction of colloidal flocks in disordered environments},
  \href{https://doi.org/10.1038/nphys3903}{\emph{Nature Physics} {\bfseries 13}
  (2017) 63}.

\bibitem{Chardac2021}
A.~Chardac, S.~Shankar, M.C.~Marchetti and D.~Bartolo, \emph{Emergence of
  dynamic vortex glasses in disordered polar active fluids},
  \href{https://doi.org/10.1073/pnas.2018218118}{\emph{Proceedings of the
  National Academy of Sciences} {\bfseries 118} (2021) }.

\bibitem{Toner2018PRL}
J.~Toner, N.~Guttenberg and Y.~Tu, \emph{Swarming in the dirt: Ordered flocks
  with quenched disorder},
  \href{https://doi.org/10.1103/PhysRevLett.121.248002}{\emph{Phys. Rev. Lett.}
  {\bfseries 121} (2018) 248002}.

\bibitem{Ballerini2008}
M.~Ballerini, N.~Cabibbo, R.~Candelier, A.~Cavagna, E.~Cisbani, I.~Giardina
  et~al., \emph{Interaction ruling animal collective behavior depends on
  topological rather than metric distance: Evidence from a field study},
  \href{https://doi.org/10.1073/pnas.0711437105}{\emph{Proceedings of the
  National Academy of Sciences} {\bfseries 105} (2008) 1232}.

\bibitem{Rahmani2021}
P.~{Rahmani}, F.~{Peruani} and P.~{Romanczuk}, \emph{{Topological flocking
  models in spatially heterogeneous environments}},
  \href{https://doi.org/10.1038/s42005-021-00708-y}{\emph{Communications
  Physics} {\bfseries 4} (2021) 206}.

\bibitem{Martin2021}
D.~Martin, H.~Chat\'e, C.~Nardini, A.~Solon, J.~Tailleur and F.~Van~Wijland,
  \emph{Fluctuation-induced phase separation in metric and topological models
  of collective motion},
  \href{https://doi.org/10.1103/PhysRevLett.126.148001}{\emph{Phys. Rev. Lett.}
  {\bfseries 126} (2021) 148001}.

\bibitem{Duan2021}
Y.~Duan, B.~Mahault, Y.-q.~Ma, X.-q.~Shi and H.~Chat\'e, \emph{Breakdown of
  ergodicity and self-averaging in polar flocks with quenched disorder},
  \href{https://doi.org/10.1103/PhysRevLett.126.178001}{\emph{Phys. Rev. Lett.}
  {\bfseries 126} (2021) 178001}.

\bibitem{chen2022}
L.~Chen, C.F.~Lee, A.~Maitra and J.~Toner, \emph{Packed swarms on dirt: two
  dimensional incompressible flocks with quenched and annealed disorder},
  \href{https://arxiv.org/abs/arXiv:2202.02865}{{\ttfamily arXiv:2202.02865}}.

\bibitem{Chen2022bis}
L.~{Chen}, C.F.~{Lee}, A.~{Maitra} and J.~{Toner}, \emph{{Incompressible polar
  active fluids with quenched disorder in dimensions $d> 2$}},
  \href{https://arxiv.org/abs/arXiv:2203.01892}{{\ttfamily arXiv:2203.01892}}.

\bibitem{Toner2018}
J.~Toner, N.~Guttenberg and Y.~Tu, \emph{Hydrodynamic theory of flocking in the
  presence of quenched disorder},
  \href{https://doi.org/10.1103/PhysRevE.98.062604}{\emph{Phys. Rev. E}
  {\bfseries 98} (2018) 062604}.

\bibitem{cardy1996}
J.~Cardy, \emph{Scaling and renormalization in statistical physics}, vol.~5,
  Cambridge university press (1996).

\bibitem{Harris1974}
A.B.~Harris, \emph{Effect of random defects on the critical behaviour of ising
  models}, \href{https://doi.org/10.1088/0022-3719/7/9/009}{\emph{Journal of
  Physics C: Solid State Physics} {\bfseries 7} (1974) 1671}.

\bibitem{Natterman1998}
T.~Nattermann, \emph{Theory of the random field ising model},  in \emph{Spin
  glasses and random fields}, pp.~277--298, World Scientific (1998).

\bibitem{Tarjus2020}
G.~Tarjus and M.~Tissier, \emph{Random-field ising and $o(n)$ models:
  theoretical description through the functional renormalization group},
  \href{https://doi.org/10.1140/epjb/e2020-100489-1}{\emph{The European
  Physical Journal B} {\bfseries 93} (2020) 50}.

\bibitem{Toner2015}
L.~{Chen}, J.~{Toner} and C.F.~{Lee}, \emph{{Critical phenomenon of the
  order-disorder transition in incompressible active fluids}},
  \href{https://doi.org/10.1088/1367-2630/17/4/042002}{\emph{New Journal of
  Physics} {\bfseries 17} (2015) 042002}.

\bibitem{attanasi2014}
A.~{Attanasi}, A.~{Cavagna}, L.~{Del Castello}, I.~{Giardina}, S.~{Melillo},
  L.~{Parisi} et~al., \emph{{Collective Behaviour without Collective Order in
  Wild Swarms of Midges}},
  \href{https://doi.org/10.1371/journal.pcbi.1003697}{\emph{PLoS Computational
  Biology} {\bfseries 10} (2014) }.

\bibitem{Mora2016}
T.~Mora, A.M.~Walczak, L.~Del~Castello, F.~Ginelli, S.~Melillo, L.~Parisi
  et~al., \emph{Local equilibrium in bird flocks},
  \href{https://doi.org/10.1038/nphys3846}{\emph{\href{https://doi.org/10.1038/nphys3846}{Nature
  Physics}} {\bfseries 12} (2016) 1153}.

\bibitem{Maitra2020}
A.~Maitra, P.~Srivastava, M.C.~Marchetti, S.~Ramaswamy and M.~Lenz,
  \emph{Swimmer suspensions on substrates: Anomalous stability and long-range
  order}, \href{https://doi.org/10.1103/PhysRevLett.124.028002}{\emph{Phys.
  Rev. Lett.} {\bfseries 124} (2020) 028002}.

\bibitem{TonerTu1998}
J.~Toner and Y.~Tu, \emph{Flocks, herds, and schools: A quantitative theory of
  flocking}, \href{https://doi.org/10.1103/PhysRevE.58.4828}{\emph{Phys. Rev.
  E} {\bfseries 58} (1998) 4828}.

\bibitem{Medina1989}
E.~Medina, T.~Hwa, M.~Kardar and Y.-C.~Zhang, \emph{Burgers equation with
  correlated noise: Renormalization-group analysis and applications to directed
  polymers and interface growth},
  \href{https://doi.org/10.1103/PhysRevA.39.3053}{\emph{Phys. Rev. A}
  {\bfseries 39} (1989) 3053}.

\bibitem{KPZ1986}
M.~Kardar, G.~Parisi and Y.-C.~Zhang, \emph{Dynamic scaling of growing
  interfaces}, \href{https://doi.org/10.1103/PhysRevLett.56.889}{\emph{Phys.
  Rev. Lett.} {\bfseries 56} (1986) 889}.

\bibitem{MSR}
P.C.~Martin, E.D.~Siggia and H.A.~Rose, \emph{Statistical dynamics of classical
  systems}, \href{https://doi.org/10.1103/PhysRevA.8.423}{\emph{Phys. Rev. A}
  {\bfseries 8} (1973) 423}.

\bibitem{Janssen1976}
H.-K.~{Janssen}, \emph{{On a Lagrangean for classical field dynamics and
  renormalization group calculations of dynamical critical properties}},
  \href{https://doi.org/10.1007/BF01316547}{\emph{Zeitschrift fur Physik B
  Condensed Matter} {\bfseries 23} (1976) 377}.

\bibitem{dominicis1976}
C.d.~Dominicis, \emph{Technics of field renormalization and dynamics of
  critical phenomena},  in \emph{J. Phys.(Paris), Colloq}, pp.~C1--247, 1976.

\bibitem{tauberbook}
U.C.~T{\"a}uber, \emph{Critical Dynamics: A Field Theory Approach to
  Equilibrium and Non-Equilibrium Scaling Behavior}, {Cambridge University
  Press}, {Cambridge, United Kingdom} (2014).

\bibitem{vasilievbook}
A.N.~Vasil'ev, \emph{The field theoretic renormalization group in critical
  behavior theory and stochastic dynamics}, {Chapman Hall/CRC}, {Boca Raton,
  FL} (2004).

\bibitem{Forster1977}
D.~Forster, D.R.~Nelson and M.J.~Stephen, \emph{Large-distance and long-time
  properties of a randomly stirred fluid},
  \href{https://doi.org/10.1103/PhysRevA.16.732}{\emph{Phys. Rev. A} {\bfseries
  16} (1977) 732}.

\bibitem{Fisher1973}
M.E.~Fisher and A.~Aharony, \emph{Dipolar interactions at ferromagnetic
  critical points},
  \href{https://doi.org/10.1103/PhysRevLett.30.559}{\emph{Phys. Rev. Lett.}
  {\bfseries 30} (1973) 559}.

\bibitem{Parisi1979}
G.~Parisi and N.~Sourlas, \emph{Random magnetic fields, supersymmetry, and
  negative dimensions},
  \href{https://doi.org/10.1103/PhysRevLett.43.744}{\emph{Phys. Rev. Lett.}
  {\bfseries 43} (1979) 744}.

\bibitem{nprg-rev}
N.~{Dupuis}, L.~{Canet}, A.~{Eichhorn}, W.~{Metzner}, J.M.~{Pawlowski},
  M.~{Tissier} et~al., \emph{{The nonperturbative functional renormalization
  group and its applications}},
  \href{https://doi.org/10.1016/j.physrep.2021.01.001}{\emph{Phys. Rep.}
  {\bfseries 910} (2021) 1}.

\bibitem{Cavagna2021}
A.~{Cavagna}, L.~{Di Carlo}, I.~{Giardina}, T.S.~{Grigera}, S.~{Melillo},
  L.~{Parisi} et~al., \emph{{Natural Swarms in $\bf 3.99$ Dimensions}},
  \href{https://arxiv.org/abs/arXiv:2107.04432}{{\ttfamily arXiv:2107.04432}}.

\bibitem{solon2022}
A.~Solon, H.~Chat{\'e}, J.~Toner and J.~Tailleur, \emph{Susceptibility of polar
  flocks to spatial anisotropy},
  \href{https://arxiv.org/abs/arXiv:2201.00704}{{\ttfamily arXiv:2201.00704}}.

\bibitem{Kardar2021}
S.~Ro, Y.~Kafri, M.~Kardar and J.~Tailleur, \emph{Disorder-induced long-ranged
  correlations in scalar active matter},
  \href{https://doi.org/10.1103/PhysRevLett.126.048003}{\emph{Phys. Rev. Lett.}
  {\bfseries 126} (2021) 048003}.

\bibitem{Elsen2018}
E.~Tjhung, C.~Nardini and M.E.~Cates, \emph{Cluster phases and bubbly phase
  separation in active fluids: Reversal of the ostwald process},
  \href{https://doi.org/10.1103/PhysRevX.8.031080}{\emph{Phys. Rev. X}
  {\bfseries 8} (2018) 031080}.

\bibitem{Xia-qing2020}
X.-q.~Shi, G.~Fausti, H.~Chat\'e, C.~Nardini and A.~Solon, \emph{Self-organized
  critical coexistence phase in repulsive active particles},
  \href{https://doi.org/10.1103/PhysRevLett.125.168001}{\emph{Phys. Rev. Lett.}
  {\bfseries 125} (2020) 168001}.

\bibitem{Ric2021}
S.~Mahdisoltani, R.~Ben Al\`{\i}~Zinati, C.~Duclut, A.~Gambassi and
  R.~Golestanian, \emph{Nonequilibrium polarity-induced chemotaxis: Emergent
  galilean symmetry and exact scaling exponents},
  \href{https://doi.org/10.1103/PhysRevResearch.3.013100}{\emph{Phys. Rev.
  Research} {\bfseries 3} (2021) 013100}.

\bibitem{Ric2022}
R.~Ben Al\`{\i}~Zinati, C.~Duclut, S.~Mahdisoltani, A.~Gambassi and
  R.~Golestanian, \emph{Stochastic dynamics of chemotactic colonies with
  logistic growth},
  \href{https://doi.org/https://doi.org/10.1209/0295-5075/ac48c9}{\emph{Europhysics
  Letters} (2022) }.

\bibitem{Wetterich1993}
C.~Wetterich, \emph{Exact evolution equation for the effective potential},
  \href{https://doi.org/https://doi.org/10.1016/0370-2693(93)90726-X}{\emph{Physics
  Letters B} {\bfseries 301} (1993) 90}.

\end{thebibliography}\endgroup
\end{document}